\documentclass[a4paper,12pt]{article}

\usepackage{amsmath}
\usepackage{amssymb}
\usepackage[dvips]{graphicx}
\usepackage[dvips]{psfrag}

\newcommand{\id}{{1\!\!1}} 
\newcommand {\beq}{\begin{equation}}
\newcommand {\eeq}{\end{equation}}
\newcommand {\bea}{\begin{eqnarray}}
\newcommand {\eea}{\end{eqnarray}}
\newcommand {\nn}{\nonumber}
\newcommand {\tr}{{\rm tr}}

\newcommand {\dd}{\mbox{d}}
\newcommand {\der}{\partial}

\newcommand {\naiveeq}{\stackrel{\rm naive}{=}}
\newcommand{\cN}{{\cal N}}
\newcommand{\cO}{{\cal O}}
\newcommand{\vev}[1]{\left\langle #1 \right\rangle}
\newcommand{\dvev}[1]{\left\langle\!\left\langle #1 \right\rangle\!\right\rangle}

\newcommand{\thth}[1]{\left. #1 \right|_{\theta\theta}}
\newcommand{\tbtb}[1]{\left. #1 \right|_{\bar{\theta}\bar{\theta}}}
\newcommand{\tttt}[1]{\left. #1 \right|_{\theta\theta\bar{\theta}\bar{\theta}}}
\newcommand{\cD}{{\cal D}}
\newcommand{\twm}{\widetilde{m}}
\newcommand{\whPhi}{\widehat{\Phi}}
\newcommand{\bq}{\bar{q}}
\newcommand{\hq}{\hat{q}}
\newcommand{\wt}{\widetilde}
\newcommand {\bA}{\tt A}
\newcommand {\bB}{\tt B}

\begin{document}
\thispagestyle{empty} \addtocounter{page}{-1}
\begin{flushright}
OIQP-08-07\\
%
\end{flushright} 
\vspace*{1cm}

\centerline{\Large \bf  Lattice Formulation of Two-Dimensional $\cN =(2,2)$ SQCD}
\vskip0.4cm
\centerline{\Large \bf with Exact Supersymmetry} 
\vspace*{2cm}
\centerline{\large Fumihiko Sugino} 
\vspace*{1.0cm}
\centerline{\it Okayama Institute for Quantum Physics} \vspace*{0.2cm}
\centerline{\it Kyoyama 1-9-1, Okayama 700-0015, Japan}
\vspace*{0.8cm}
\centerline{\tt fumihiko\_sugino@pref.okayama.lg.jp} 
\vskip2cm
\centerline{\bf Abstract}
\vspace*{0.3cm}
{\small 

We construct a lattice model for two-dimensional ${\cal N}=(2,2)$ supersymmetric QCD (SQCD), 
with the matter multiplets belonging to the fundamental or anti-fundamental representation of the 
gauge group ${\rm U}(N)$ or ${\rm SU}(N)$. 
The construction is based on the topological field theory (twisted supercharge) formulation 
and exactly preserves one supercharge 
along the line of the papers~\cite{sugino}--\cite{sugino4} for pure supersymmetric Yang-Mills theories. 
In order to avoid the species doublers of the matter multiplets, we introduce the Wilson terms and the model 
is defined for the case of the number of the fundamental matters ($n_+$) equal to 
that of the anti-fundamental matters ($n_-$). 
If some of the matter multiplets decouple from the theory by sending the corresponding anti-holomorphic twisted masses 
to the infinity, we can analyze the general $n_+\neq n_-$ case, although the lattice model is defined for $n_+ =n_-$. 
By computing the anomaly of the ${\rm U}(1)_A$ R-symmetry in the lattice perturbation, we see that 
the decoupling is achieved and the anomaly for $n_+\neq n_-$ is correctly obtained.       
}
\vspace*{1.1cm}



\newpage


\section{Introduction}
Lattice formulation of quantum field theory has been the most solid method to give its constructive definition 
and to explore its nonperturbative properties. 
Wilson's lattice gauge theory has been demonstrating it as the most typical and well-known example. 
In general, it is not possible to realize all the symmetries possessed by the target continuum theory in the lattice formulation.  
Thus, in constructing the target theory from its lattice formulation, in the continuum limit 
we often have to tune some coupling constants against the corresponding relevant operators radiatively generated, 
to recover the symmetries not realized at the lattice level. 
It is desirable to start with the lattice theory realizing more symmetries of those possessed by the target theory, so that 
the relevant operators to be tuned are fewer. 
Also, from the theoretical point of view, it is intriguing to seek novel ultra-violet completions which possess symmetries 
never realized so far. The Ginsparg-Wilson formulation for the chiral symmetry~\cite{ginsparg-wilson} is such a well-known example.  
 
Since supersymmetric gauge theory is one of the promising candidates that describe the physics beyond the standard model, 
it is important to construct its lattice formulation to proceed the nonperturbative investigation from the first principle. 
However, there has been a notorious difficulty on the reconciliation of supersymmetry 
with the lattice structure~\cite{curci-veneziano,kato-ss,bergner}\footnote{
Although there is an attempt to overcome the difficulty by deforming the Leibniz rule on the lattice~\cite{kawamoto,nagata}, 
it seems necessary to be elaborated further~\cite{bruckmann}.}. 
At present, some lattice models realizing a part of the supersymmetries of the target theory have been constructed\footnote{
For a recent review, see \cite{giedt}.}.  
For example, see \cite{sakai-sakamoto,catterall,catterall4,kikukawa-nakayama} for the field theories without gauge symmetry, 
\cite{kaplan,harada-pinsky,sugino,sugino2,sugino3,catterall2,sugino4,tsuchiya} for pure supersymmetric Yang-Mills (SYM) models\footnote{
Ref.~\cite{ohta-takimi} discusses on observables of the topological field theory on the lattice.}, 
and \cite{endre-kaplan,matsuura} for the two-dimensional SYM coupled with matter fields. 
Moreover, the numerical simulations of the constructed lattice two-dimensional SYM models have been 
done in~\cite{catterall3,suzuki,kanamori-ss}. 
In particular, Refs.~\cite{kanamori-ss} have presented a framework of observing 
the dynamical supersymmetry breaking for a general lattice theory possessing at least one exact supercharge. 

In this paper, we construct a lattice theory for two-dimensional $\cN=(2,2)$ supersymmetric QCD (SQCD) with 
matter multiplets belonging to the fundamental or anti-fundamental representation of 
the gauge group $G={\rm U}(N)$ or ${\rm SU}(N)$. 
In the same manner as the previous work for SYM theories~\cite{sugino,sugino2,sugino3,sugino4}, 
our construction is based on the topological field theory (or twisted supercharge) formulation of the 
target supersymmetric theory, differently from \cite{endre-kaplan,matsuura} based on the idea of the deconstruction\footnote{
Some relations among the deconstruction models and the topological field theory construction are 
discussed in~\cite{unsal-takimi,damgaard-m}.}. 
The lattice gauge fields are represented as compact link variables, and one of the supercharges of the target theory 
is exactly preserved at the lattice level. 
 
This paper is organized as follows. 
In the next section, we explain the target continuum theory, two-dimensional $\cN =(2,2)$ SQCD 
with $n_+$ fundamental and $n_-$ anti-fundamental matter multiplets. 
We can introduce general superpotentials and the twisted masses (furthermore the Fayet-Iliopoulos (FI) term 
and the topological $\vartheta$-term for $G={\rm U}(N)$). 
The action is supersymmetric and expressed as the $Q$-exact form (except for the topological $\vartheta$-term). 
$Q$ is a linear combination of the four supercharges of the target theory obtained by the topological twist. 
In the presence of the twisted masses, $Q$ is nilpotent up to the combination of an infinitesimal gauge transformation and infinitesimal flavor rotations, whose transformation parameters are the 
Higgs scalar $\phi$ and the holomorphic twisted masses, respectively. 
In section~\ref{sec:lat_2DSQCD}, we construct the lattice action with the supersymmetry $Q$ exactly preserved. 
In order to avoid the species doublers of the matter multiplets, we introduce the Wilson terms and the model is defined 
in the case $n_+ = n_-(\equiv n)$. Then, the flavor symmetry reduces from ${\rm U}(1)^n \times {\rm U}(1)^n$ 
to its diagonal subgroup ${\rm U}(1)^n$. 
The $Q$-invariance of our lattice action is guaranteed, 
when the flavor rotation generated by $Q^2$ falls into the subgroup ${\rm U}(1)^n$. 
Thus, we are forced to focus on the case that the holomorphic twisted masses of the fundamentals and anti-fundamentals 
of the flavor $I$ are equal $\twm_{+I}=\twm_{-I}(\equiv \twm_I)$ with $I=1, \cdots, n$.  
In section~\ref{sec:anomaly}, we analyze the anomaly of the ${\rm U}(1)_A$ R-symmetry in both of the continuum and lattice cases. 
Although the lattice action is applicable only to the case $n_+ =n_-$, it will be possible to obtain the physical consequences 
for the general case $n_+ \neq n_-$, by sending some of the anti-holomorphic twisted masses to the infinity 
intending to decouple the corresponding matters from the theory. 
Actually, we see that the decoupling is achieved and the ${\rm U}(1)_A$ anomaly for $n_+\neq n_-$ is correctly obtained 
by perturbative calculation using our lattice action. 
The summary of the results obtained so far and the discussion on future subjects are presented in section~\ref{sec:summary}. 
In appendix~\ref{app:cont(1+1)DSQCD}, to clarify the notation, we explicitly derive the $(1+1)$-dimensional $\cN =(2,2)$ 
SQCD action by the dimensional reduction from $\cN=1$ SQCD in $3+1$ dimensions. 
Appendix~\ref{app:lat_pert} is devoted to details on the lattice perturbative computation of the ${\rm U}(1)_A$ anomaly.

\section{Two-Dimensional Continuum ${\cal N}= (2,2)$ SQCD} 
\label{sec:cont_2DSQCD}
\setcounter{equation}{0}
$\cN =(2,2)$ SQCD in $1+1$ dimensions is derived from the $(3+1)$-dimensional $\cN =1$ SQCD 
by the dimensional reduction. 
The field contents are the dimensional reduction of the four-dimensional vector multiplet $V$, 
$n_+$ chiral multiplets belonging to the fundamental representation 
$\Phi_{+I} =(\phi_{+I}, \psi_{+I}, F_{+I})$ $(I=1, \cdots, n_+)$, and 
$n_-$ chiral multiplets belonging to the anti-fundamental representation 
$\Phi_{-I'} =(\phi_{-I'}, \psi_{-I'}, F_{-I'})$ $(I' =1, \cdots, n_-)$. 
After the dimensional reduction, $V$ contains the gauge fields $A_{\mu}$, the Higgs scalars $\phi, \bar{\phi}$, 
the gaugino fields $\lambda, \bar{\lambda}$, and the auxiliary field $D$.   
The detail is explained in appendix~\ref{app:cont(1+1)DSQCD}. 

To develop the corresponding lattice formulation, we consider the theory in Euclidean two dimensions, 
which is obtained from (\ref{2dSQCD_Min}) by the Wick rotation 
\beq
x^0 \to -ix_0, \qquad A_0 \to iA_0. 
\eeq
The result is 
\bea
S^{(E)}_{\rm 2DSQCD} & = & S^{(E)}_{\rm 2DSYM} + S^{(E)}_{{\rm mat}, +} + S^{(E)}_{{\rm mat}, -}, 
\label{2dSQCD_action}\\
S^{(E)}_{\rm 2DSYM} & = & \frac{1}{g^2} \int \dd^2x \, \tr \left(\frac12
F_{\mu\nu}F_{\mu\nu} 
+\cD_\mu\phi\cD_\mu\bar{\phi} 
+\frac14[\phi, \bar{\phi}]^2 -D^2 \right. \nn \\
 & & \left. \hspace{2.5cm} +4\bar{\lambda}_R\cD_z\lambda_R + 4\bar{\lambda}_L\cD_{\bar{z}}\lambda_L 
+2\bar{\lambda}_R[\bar{\phi}, \lambda_L] +2\bar{\lambda}_L [\phi, \lambda_R] \frac{}{} \right), \nn \\
S^{(E)}_{{\rm mat}, +} & = & \int \dd^2x \sum_{I=1}^{n_+}\left[\cD_\mu\phi_{+I}^\dagger\cD_\mu\phi_{+I} 
+\frac12\phi_{+I}^\dagger\{\phi, \bar{\phi}\}\phi_{+I} -F_{+I}^\dagger
F_{+I}
-\phi_{+I}^\dagger D\phi_{+I} \right. \nn \\
 & & \hspace{1.5cm} +2\bar{\psi}_{+IR}\cD_z\psi_{+IR} + 2\bar{\psi}_{+IL}\cD_{\bar{z}}\psi_{+IL} 
+\bar{\psi}_{+IR} \, \bar{\phi} \, \psi_{+IL} +\bar{\psi}_{+IL} \, \phi \, \psi_{+IR} \nn \\
 & & \hspace{1.5cm} \left. -i\sqrt{2}\left(\phi_{+I}^\dagger(\lambda_L\psi_{+IR}-\lambda_R\psi_{+IL}) 
 +(-\bar{\psi}_{+IR}\bar{\lambda}_L+\bar{\psi}_{+IL}\bar{\lambda}_R)\phi_{+I}\right)\right], \nn \\
S^{(E)}_{{\rm mat}, -} & = &  \int \dd^2x
\sum_{I'=1}^{n_-}\left[\cD_\mu\phi_{-I'}\cD_\mu\phi_{-I'}^\dagger 
+\frac12\phi_{-I'}\{\phi, \bar{\phi}\}\phi_{-I'}^\dagger 
-F_{-I'}F_{-I'}^\dagger +\phi_{-I'} D\phi_{-I'}^\dagger \right. \nn \\
 & & \hspace{1.5cm} +2\psi_{-I'R}\cD_z\bar{\psi}_{-I'R} +2\psi_{-I'L}\cD_{\bar{z}}\bar{\psi}_{-I'L} 
 +\psi_{-I'L} \, \bar{\phi} \, \bar{\psi}_{-I'R} +\psi_{-I'R} \, \phi \, \bar{\psi}_{-I'L} \nn \\
 & & \hspace{1.5cm} \left. -i\sqrt{2}\left((-\psi_{-I'L}\lambda_R +\psi_{-I'R}\lambda_L)\phi_{-I'}^\dagger 
 +\phi_{-I'} (\bar{\lambda}_R\bar{\psi}_{-I'L} -\bar{\lambda}_L\bar{\psi}_{-I'R})\right) \right], \nn
\eea
where $\cD_z = \frac12(\cD_0-i\cD_1)$, $\cD_{\bar{z}}=\frac12 (\cD_0+i\cD_1)$, 
and the spinor indices $R$, $L$ are used instead of 1, 2, respectively. 
After the Wick rotation, the contours of the auxiliary fields in the path-integral are chosen 
to give the convergent result\footnote{
Or equivalently, treating $F_{+I}$ and $F_{+I}^\dagger$ ($F_{-I'}$ and $F_{-I'}^\dagger$) as independent variables, 
we further rotate as $D \to iD$, $F_{+I} \to iF_{+I}$ and $F_{+I}^\dagger \to iF_{+I}^\dagger$ 
($F_{-I'} \to iF_{-I'}$ and $F_{-I'}^\dagger \to iF_{-I'}^\dagger$).}.  

The supersymmetry of the action is given by the result of the Wick rotation 
of (\ref{2dsusy_SYM}), (\ref{2dsusy_mat+}) and (\ref{2dsusy_mat-}). 
Also, the classical action has the two R-symmetries ${\rm U(1)}_V$ and ${\rm U(1)}_A$. 
The origin of the ${\rm U(1)}_V$ is the ${\rm U(1)}_R$ symmetry in the four-dimensional $\cN =1$ theory, 
while that of the ${\rm U(1)}_A$ is the rotational symmetry in the directions $(x'^1, x'^2)$ to be dimensionally reduced. 
The $({\rm U(1)}_V, {\rm U(1)}_A)$ charges are assigned 
for the supercoordinates $\theta, \bar{\theta}$, the supercharges $Q, \bar{Q}$ and the field contents as 
\bea
& & \theta_L :(1,1), \quad \theta_R:(1,-1), \quad \bar{\theta}_L :(-1,-1), \quad \bar{\theta}_R :(-1, 1) \nn \\
& & Q_L:(-1, 1), \quad Q_R :(-1,-1), \quad \bar{Q}_L:(1,-1), \quad \bar{Q}_R:(1,1) \nn \\
& & \phi :(0,2), \quad \bar{\phi}:(0,-2), \quad \lambda_L:(1,1), \quad \lambda_R:(1,-1), \quad \bar{\lambda}_L:(-1,-1), \quad 
\bar{\lambda}_R:(-1,1) \nn \\
& & \psi_{+I L}:(-1,1), \quad \psi_{+I R}:(-1,-1), \quad F_{+I}:(-2,0), \nn \\
 & & \bar{\psi}_{+IL}:(1,-1), \quad \bar{\psi}_{+I R}:(1,1), \quad F_{+I}^\dagger :(2,0), \nn \\
& & \psi_{-I' L}:(-1,1), \quad \psi_{-I' R}:(-1,-1), \quad F_{-I'}:(-2,0), \nn \\
 & & \bar{\psi}_{-I' L}:(1,-1), \quad \bar{\psi}_{-I' R}:(1,1), \quad F_{-I'}^\dagger :(2,0)
\eea
with the other fields neutral for both ${\rm U(1)}$'s.  

Renaming the variables as\footnote{This notation is based on the representation under the twisted ${\rm U(1)}$ group 
-- the diagonal sum of the two-dimensional rotational group and ${\rm U(1)}_V$, which corresponds to 
the $A$ model twist in Ref.~\cite{witten2}. }   
\bea
& & Q \equiv -\frac{1}{\sqrt{2}} (Q_L + \bar{Q}_R), \nn \\
& & \psi_0 \equiv \frac{1}{\sqrt{2}}(\lambda_L +\bar{\lambda}_R), \qquad \psi_1 \equiv \frac{i}{\sqrt{2}}(\lambda_L-\bar{\lambda}_R), 
\nn \\
& & \chi \equiv \frac{1}{\sqrt{2}}(\lambda_R-\bar{\lambda}_L), \qquad \eta \equiv -i\sqrt{2}(\lambda_R +\bar{\lambda}_L), 
\eea
the transformation rule under the supersymmetry $Q$ is given by  
\bea
 & & QA_\mu =\psi_\mu, \qquad Q\psi_\mu = i\cD_\mu\phi, \nn \\
 & & Q\phi =0, \nn \\
 & & Q\bar{\phi} =\eta, \qquad Q\eta = [\phi, \bar{\phi}], \nn \\
 & & Q\chi = iD +iF_{01}, \qquad QD = -QF_{01} -i[\phi, \chi], 
\label{Qsusy_cont_SYM} 
\eea
\bea
 & & Q\phi_{+I} = -\psi_{+IL}, \qquad Q\psi_{+IL} = -\phi \, \phi_{+I}, \qquad Q\psi_{+IR} = (\cD_0 +i\cD_1)\phi_{+I} + F_{+I}, \nn \\
 & & QF_{+I} = (\cD_0 +i\cD_1)\psi_{+IL} +\phi \, \psi_{+IR} -i(\psi_0+i\psi_1)\phi_{+I}, \nn \\
 & & Q\phi_{+I}^\dagger = -\bar{\psi}_{+IR}, \qquad Q\bar{\psi}_{+IR} = \phi_{+I}^\dagger \, \phi, \qquad 
Q\bar{\psi}_{+IL} = (\cD_0-i\cD_1)\phi_{+I}^\dagger + F_{+I}^\dagger, \nn \\
 & & QF_{+I}^\dagger = (\cD_0-i\cD_1)\bar{\psi}_{+IR} -\bar{\psi}_{+IL} \, \phi +i\phi_{+I}^\dagger (\psi_0-i\psi_1), 
\label{Qsusy_cont_mat+}
\eea
\bea 
 & & Q\phi_{-I'} = -\psi_{-I'L}, \qquad Q\psi_{-I'L} = -\phi_{-I'} \, \phi, \qquad 
Q\psi_{-I'R} = (\cD_0+i\cD_1)\phi_{-I'} +F_{-I'}, \nn \\
 & & QF_{-I'} = (\cD_0+i\cD_1)\psi_{-I'L} -\psi_{-I'R} \, \phi +i\phi_{-I'}(\psi_0+i\psi_1), \nn \\
 & & Q\phi_{-I'}^\dagger = -\bar{\psi}_{-I'R}, \qquad Q\bar{\psi}_{-I'R} =-\phi \, \phi_{-I'}^\dagger, \qquad 
Q\bar{\psi}_{-I'L} = (\cD_0 -i\cD_1)\phi_{-I'}^\dagger + F_{-I'}^\dagger, \nn \\
 & & QF_{-I'}^\dagger = (\cD_0 -i\cD_1)\bar{\psi}_{-I'R} + \phi \, \bar{\psi}_{-I'L} -i(\psi_0-i\psi_1)\phi_{-I'}^\dagger. 
\label{Qsusy_cont_mat-}
\eea
It is easily seen that $Q$ is nilpotent up to the infinitesimal gauge transformation with the (complexified) parameter $\phi$. 
The action (\ref{2dSQCD_action}) can be expressed as the $Q$-exact form: 
\bea
S^{(E)}_{\rm SYM} & = & Q\frac{1}{g^2} \int \dd^2x \, \tr \left[-i\chi (F_{01}-D) +\frac14 \eta [\phi, \bar{\phi}] 
-i\psi_\mu\cD_\mu\bar{\phi}\right], 
\label{S_cont_SYM}\\
S^{(E)}_{{\rm mat}, +} & = & Q \int\dd^2x \sum_{I=1}^{n_+}\frac12\left[\frac{}{} \bar{\psi}_{+IL}\left((\cD_0+i\cD_1)\phi_{+I} -F_{+I}\right) 
\right. \nn \\
 & & \hspace{2.5cm} +\left((\cD_0 -i\cD_1)\phi_{+I}^\dagger -F_{+I}^\dagger\right)\psi_{+IR} \nn \\
 & & \hspace{2.5cm} \left. +\bar{\psi}_{+IR} \, \bar{\phi} \, \phi_{+I} -\phi_{+I}^\dagger \, \bar{\phi} \, \psi_{+IL} 
+2i\phi_{+I}^\dagger\chi \phi_{+I}\right], \\
S^{(E)}_{{\rm mat}, -} & = &  Q \int\dd^2x \sum_{I'=1}^{n_-}\frac12\left[ \frac{}{} \left((\cD_0+i\cD_1)\phi_{-I'} -F_{-I'}\right)\bar{\psi}_{-I'L} 
\right. \nn \\
 & & \hspace{2.5cm} +\psi_{-I'R}\left((\cD_0-i\cD_1)\phi_{-I'}^\dagger -F_{-I'}^\dagger\right) \nn \\
 & & \hspace{2.5cm} \left. -\phi_{-I'} \, \bar{\phi} \, \bar{\psi}_{-I'R} +\psi_{-I'L} \, \bar{\phi} \, \phi_{-I'}^\dagger 
-2i\phi_{-I'}\chi\phi_{-I'}^\dagger\right]. 
\eea
In these formulas, $Q$ acts to the gauge invariant expressions, from which the $Q$ invariance of the actions follows. 
By rewriting the auxiliary field as $H \equiv iD +iF_{01}$, the SYM sector exactly coincides to the corresponding 
SYM action discussed in~\cite{sugino,sugino2}.  
    
\subsection{Superpotentials, Mass Terms, FI and $\vartheta$-Terms}
We can introduce interactions among the matter fields $\Phi_{+I}, \Phi_{-I'}$ in the form of 
the dimensional reduction of the superpotentials in $3+1$ dimensions: 
\beq
\int \dd^4x' \, \left[\thth{W(\Phi_+, \Phi_-) \frac{}{} } 
+ \tbtb{\bar{W}(\Phi_+^\dagger, \Phi_-^\dagger) \frac{}{} }\right], 
\eeq
where the gauge and flavor indices of the fields are appropriately contracted. 

Here, two kinds of mass terms can be introduced to the matters. One is the complex mass terms, which are included 
in the superpotentials. The fermion part is 
\beq
\sum_{I,I'} \left[m_{II'}\left(\psi_{-I'L} \, \psi_{+IR} -\psi_{-I'R} \, \psi_{+IL}\right) 
+m_{I'I}^* \left(\bar{\psi}_{+IR} \, \bar{\psi}_{-I'L} -\bar{\psi}_{+IL} \, \bar{\psi}_{-I'R}\right) \right].
\eeq
The complex masses preserve ${\rm U}(1)_A$, but not ${\rm U}(1)_V$. 
Then, ${\rm U}(1)_V$ combined with ${\rm U}(1)_a$, which is a ${\rm U}(1)$-subgroup of the flavor rotation: 
\bea
{\rm U}(1)_a: \qquad \Phi_{+I} \to e^{i\alpha} \, \Phi_{+I}, & & \Phi_{-I'} \to e^{i\alpha} \, \Phi_{-I'}, \nn \\
\Phi_{+I}^\dagger \to e^{-i\alpha} \, \Phi_{+I}^\dagger, & & \Phi_{-I'}^\dagger \to e^{-i\alpha} \, \Phi_{-I'}^\dagger, 
\qquad (\alpha\in {\bf R})
\eea
can become a symmetry. 

The other is the twisted mass terms, which are not included in the superpotentials~\cite{hanany-hori}. 
They can be introduced by gauging the ${\rm U}(1)^{n_+}\times {\rm U}(1)^{n_-}$ of the flavor symmetry 
and fixing the corresponding vector superfields to the background values as 
\bea
\sum_{I=1}^{n_+}\Phi_{+I}^\dagger \, e^{V} \, \Phi_{+I} & \to & 
\sum_{I=1}^{n_+}\Phi_{+I}^\dagger \, e^{V -\widetilde{V}_{+I}} \, \Phi_{+I}, \nn \\
\sum_{I'=1}^{n_-} \Phi_{-I'} \, e^{-V} \, \Phi_{-I'}^\dagger & \to & 
\sum_{I'=1}^{n_-} \Phi_{-I'} \, e^{-V + \widetilde{V}_{-I'}} \, \Phi_{-I'}^\dagger
\eea
with 
\bea
\widetilde{V}_{+I} & \equiv & 2\theta_R\bar{\theta}_L \, \twm_{+I} +2\theta_L\bar{\theta}_R \,\twm_{+I}^* \, , \nn \\
\widetilde{V}_{-I'} & \equiv & 2\theta_R\bar{\theta}_L \, \twm_{-I'} +2\theta_L\bar{\theta}_R \, \twm_{-I'}^* \, . 
\eea
They give the following mass terms to the fermions:  
\beq
\sum_{I=1}^{n_+}\left(\twm_{+I} \, \bar{\psi}_{+IL} \, \psi_{+IR} +\twm_{+I}^* \, \bar{\psi}_{+IR} \, \psi_{+IL}\right) 
+\sum_{I'=1}^{n_-}\left(\twm_{-I'} \, \psi_{-I'R} \, \bar{\psi}_{-I'L} +\twm_{-I'}^* \, \psi_{-I'L} \, \bar{\psi}_{-I'R}\right). 
\eeq
The twisted masses preserve ${\rm U}(1)_V$, but not ${\rm U}(1)_A$. 
It should be noted that the flavor symmetry of the matter-part action is broken from ${\rm U}(n_+)\times {\rm U}(n_-)$ 
to ${\rm U}(1)^{n_+} \times {\rm U}(1)^{n_-}$ by introducing generic twisted masses. 

Because ${\rm U}(1)_A$ transforms the left-handed fermions and the right-handed fermions differently, 
it can be anomalous at the quantum level. We will discuss it in section~\ref{sec:anomaly}. 

In the presence of the twisted masses, the supersymmetry transformations for the matters are deformed. 
(For the explicit form, see (\ref{2dsusy_mat+_twm}), (\ref{2dsusy_mat-_twm}).) 
In particular, the $Q$ transformation becomes
\bea
 & & Q\phi_{+I} = -\psi_{+IL}, \qquad Q\psi_{+IL} = -(\phi-\twm_{+I})\phi_{+I}, \nn \\
 & & Q\psi_{+IR} = (\cD_0 +i\cD_1)\phi_{+I} + F_{+I}, \nn \\
 & & QF_{+I} = (\cD_0 +i\cD_1)\psi_{+IL} +(\phi-\twm_{+I})\psi_{+IR} -i(\psi_0+i\psi_1)\phi_{+I}, \nn \\
 & & Q\phi_{+I}^\dagger = -\bar{\psi}_{+IR}, \qquad Q\bar{\psi}_{+IR} = \phi_{+I}^\dagger (\phi-\twm_{+I}), \nn \\
 & & Q\bar{\psi}_{+IL} = (\cD_0-i\cD_1)\phi_{+I}^\dagger + F_{+I}^\dagger, \nn \\
 & & QF_{+I}^\dagger = (\cD_0-i\cD_1)\bar{\psi}_{+IR} -\bar{\psi}_{+IL}(\phi-\twm_{+I}) +i\phi_{+I}^\dagger (\psi_0-i\psi_1), 
\label{Qsusy_cont_mat+_twm}
\eea
\bea 
 & & Q\phi_{-I'} = -\psi_{-I'L}, \qquad Q\psi_{-I'L} = -\phi_{-I'}(\phi-\twm_{-I'}), \nn \\
 & & Q\psi_{-I'R} = (\cD_0+i\cD_1)\phi_{-I'} +F_{-I'}, \nn \\
 & & QF_{-I'} = (\cD_0+i\cD_1)\psi_{-I'L} -\psi_{-I'R}(\phi-\twm_{-I'}) +i\phi_{-I'}(\psi_0+i\psi_1), \nn \\
 & & Q\phi_{-I'}^\dagger = -\bar{\psi}_{-I'R}, \qquad Q\bar{\psi}_{-I'R} =-(\phi-\twm_{-I'})\phi_{-I'}^\dagger, \nn \\
 & & Q\bar{\psi}_{-I'L} = (\cD_0 -i\cD_1)\phi_{-I'}^\dagger + F_{-I'}^\dagger, \nn \\
 & & QF_{-I'}^\dagger = (\cD_0 -i\cD_1)\bar{\psi}_{-I'R} + (\phi-\twm_{-I'})\bar{\psi}_{-I'L} -i(\psi_0-i\psi_1)\phi_{-I'}^\dagger. 
\label{Qsusy_cont_mat-_twm}
\eea
Then, $Q$ is nilpotent up to the combination of the infinitesimal gauge transformation with the (complexified) parameter $\phi$ and 
the infinitesimal flavor rotations with the (complexified) parameters $\twm_{+I}, \twm_{-I'}$ acting as 
\bea
\delta\Phi_{+I} = -\twm_{+I}\Phi_{+I}, &  & \delta\Phi_{+I}^\dagger = \twm_{+I}\Phi_{+I}^\dagger, \nn \\
\delta\Phi_{-I'} = \twm_{-I'}\Phi_{-I'}, & & \delta\Phi_{-I'}^\dagger =-\twm_{-I'}\Phi_{-I'}^\dagger. 
\eea
 
The matter-part actions, the Wick rotation of (\ref{2d_S_mat_twm+}) and (\ref{2d_S_mat_twm-}), 
can be written as the $Q$-exact form: 
\bea
S^{(E)}_{{\rm mat}, +\twm} & = & Q \int\dd^2x \sum_{I=1}^{n_+}\frac12\left[\frac{}{} 
\bar{\psi}_{+IL}\left((\cD_0+i\cD_1)\phi_{+I} -F_{+I}\right) \right. \nn \\
 & & \hspace{2.5cm} +\left((\cD_0 -i\cD_1)\phi_{+I}^\dagger -F_{+I}^\dagger\right)\psi_{+IR} \nn \\
 & & \left. +\bar{\psi}_{+IR}(\bar{\phi}-\twm_{+I}^*)\phi_{+I} 
-\phi_{+I}^\dagger(\bar{\phi}-\twm_{+I}^*)\psi_{+IL} 
+2i\phi_{+I}^\dagger\chi \phi_{+I}\right], 
\label{S_cont_mat+}\\
S^{(E)}_{{\rm mat}, -\twm} & = &  Q \int\dd^2x \sum_{I'=1}^{n_-}\frac12\left[\frac{}{} 
\left((\cD_0+i\cD_1)\phi_{-I'} -F_{-I'}\right)\bar{\psi}_{-I'L} \right. \nn \\
 & & \hspace{2.5cm} +\psi_{-I'R}\left((\cD_0-i\cD_1)\phi_{-I'}^\dagger -F_{-I'}^\dagger\right) \nn \\
 & & \left. -\phi_{-I'}(\bar{\phi}-\twm_{-I'}^*)\bar{\psi}_{-I'R} 
+\psi_{-I'L}(\bar{\phi}-\twm_{-I'}^*)\phi_{-I'}^\dagger 
-2i\phi_{-I'}\chi\phi_{-I'}^\dagger\right]. 
\label{S_cont_mat-}
\eea
In the above, $Q$ acts to the gauge invariant expressions possessing the flavor symmetry ${\rm U}(1)^{n_+}\times {\rm U}(1)^{n_-}$, 
which shows the $Q$ invariance of the actions. 

Also, the superpotential terms can be expressed as the $Q$-exact form: 
\bea
S^{(E)}_{\rm pot} & = & Q \int \dd^2x \sum_{i=1}^N\left[\sum_{I=1}^{n_+}\left(-\frac{\der W(\phi_+, \phi_-)}{\der\phi_{+Ii}}\psi_{+IRi} 
-\bar{\psi}_{+ILi}\frac{\der \bar{W}(\phi_+^\dagger, \phi_-^\dagger)}{\der \phi_{+Ii}^*} \right)\right. \nn \\
 & & \hspace{2cm} \left. +\sum_{I'=1}^{n_-}\left(-\psi_{-I'Ri}\frac{\der W(\phi_+, \phi_-)}{\der \phi_{-I'i}}
-\frac{\der \bar{W}(\phi_+^\dagger, \phi_-^\dagger)}{\der \phi_{-I'i}^*}\bar{\psi}_{-I'Li} \right)\right], 
\label{S_cont_pot} 
 \eea
where we wrote the gauge index $i(=1, \cdots, N)$ explicitly.  

For the case $G={\rm U}(N)$, the FI and $\vartheta$-terms can be introduced to the action:
\bea
S^{(E)}_{{\rm FI}, \, \vartheta} & = & \int \dd^2x \, \tr \left(\kappa D -i\frac{\vartheta}{2\pi} F_{01}\right) \nn \\
 & = &  Q \kappa \int \dd^2x \,\tr \left(-i\chi\right) 
-i \frac{\vartheta-2\pi i\kappa}{2\pi} \int \dd^2x \,\tr \, F_{01}
\label{FI_theta_cont}
\eea
with $\kappa$ being the FI parameter. 
The second term in the r.h.s. is a topological term, and thus $Q$-invariant. 
The first term yields the $\vartheta$-term with the imaginary value $\vartheta =2\pi i \kappa$, that is 
compensated by the second term. 
The $Q$-exact action gives the imaginary valued $\vartheta$-term, which is common to the four-dimensional case~\cite{witten}.

\setcounter{equation}{0}
\section{Two-Dimensional Lattice $\cN =(2,2)$ SQCD}
\label{sec:lat_2DSQCD}
In this section, we latticize the continuum theory discussed in the previous section with realizing 
the $Q$-supersymmetry exactly. 
The lattice is the two-dimensional square lattice with the spacing $a$, the sites of which are labeled by $x \in {\bf Z}^2$. 
The gauge field $A_{\mu}(x)$ is promoted to the variable $U_\mu(x) = e^{iaA_{\mu}(x)}$ on the link $(x, x+\hat{\mu})$. 
All the other fields are distributed on the lattice sites.

\subsection{SYM Part of Lattice Theory}
The supersymmetry transformation for the SYM fields (\ref{Qsusy_cont_SYM}) can be realized on the lattice as 
\bea
 & & QU_{\mu}(x) = i\psi_{\mu}(x) U_{\mu}(x), \qquad QU_{\mu}(x)^{-1} = -iU_{\mu}(x)^{-1} \psi_{\mu}(x), \nn \\
 & & Q\psi_{\mu}(x) = i\psi_{\mu}(x)\psi_{\mu}(x) -i\left(\phi(x) -U_{\mu}(x)\phi(x+\hat{\mu}) U_{\mu}(x)^{-1}\right), \nn \\
 & & Q\phi(x) = 0, \nn \\
 & & Q\bar{\phi}(x) = \eta(x), \qquad Q\eta(x) = [\phi(x), \bar{\phi}(x)], \nn \\
 & & Q\chi(x) = iD(x) + \frac{i}{2}\whPhi(x), \qquad 
 QD(x) = -\frac12Q\whPhi(x) -i[\phi(x), \chi(x)]\, ,
\label{Qsusy_lat_SYM} 
\eea
where $\whPhi(x)$ is a lattice counterpart of $2F_{01}(x)$ defined by 
\bea
 & & \Phi(x) = -i(U_{01}(x)-U_{10}(x)), \qquad 
U_{\mu\nu}(x) \equiv U_{\mu}(x)U_{\nu}(x+\hat{\nu})U_{\mu}(x+\hat{\nu})^{-1}U_{\nu}(x)^{-1}, \nn \\
 & & \whPhi(x) \equiv \frac{\Phi(x)}{1-\frac{1}{\epsilon^2}||1-U_{01}(x)||^2}. 
\eea
The norm of an arbitrary complex matrix $A$ is defined as $||A||\equiv \sqrt{\tr (AA^\dagger)}$, and 
$\epsilon$ is a constant chosen as 
\beq
0<\epsilon < 2  \qquad \text{for} \quad G={\rm U}(N). 
\label{admissibility_U(N)}
\eeq
In the case $G={\rm SU}(N)$, here and in what follows, $\whPhi(x)$ is understood to be replaced with its traceless part: 
\beq
\whPhi_{\rm TL}(x) \equiv \whPhi(x) - \frac{1}{N}\left(\tr \,\whPhi(x)\right)\id_N, 
\eeq
and $\epsilon$ is chosen as 
\bea
 & & 0<\epsilon <2\sqrt{2} \hspace{2.7cm}\mbox{for}\quad G={\rm SU}(2), {\rm SU}(3), {\rm SU}(4), \nn \\
 & & 0<\epsilon <2\sqrt{N}\sin\left(\frac{\pi}{N}\right) \qquad \mbox{for}\quad G={\rm SU}(N) \quad (N\geq 5). 
\eea  
The transformation (\ref{Qsusy_lat_SYM}) is defined for the lattice gauge fields satisfying the admissibility condition: 
\beq
||1-U_{01}(x)||<\epsilon, 
\label{admissibility}
\eeq
and $Q$ is nilpotent up to the infinitesimal gauge transformation with the parameter $\phi(x)$ on the lattice. 

We can construct the $Q$-invariant lattice action as the $Q$-exact form: 
\bea
S^{\rm LAT}_{\rm 2DSYM} & = & Q\frac{1}{g_0^2} \sum_x\tr\left[\chi(x)\left(-\frac{i}{2}\whPhi(x) +iD(x)\right) 
+\frac14\eta(x)[\phi(x), \bar{\phi}(x)] \right. \nn \\
 & & \hspace{2cm} \left. +i\sum_{\mu=0}^1\psi_\mu(x)\left(\bar{\phi}(x)-U_\mu(x)\bar{\phi}(x+\hat{\mu})U_\mu(x)^{-1}\right)\right] 
\label{S_lat_SYM1} 
\eea
for the admissible gauge fields satisfying (\ref{admissibility}) for $\forall x$, and 
\beq
 S^{\rm LAT}_{\rm 2DSYM} =+\infty \qquad \mbox{otherwises}.
\label{S_lat_SYM2} 
\eeq
It is straightforward to see that the $Q$-action in the r.h.s. of (\ref{S_lat_SYM1}) gives 
\bea
S^{\rm LAT}_{\rm 2DSYM} & = & \frac{1}{g_0^2}\sum_x\tr\left[\frac14\whPhi(x)^2 +i\chi(x)Q\whPhi(x) -D(x)^2 \right. \nn \\
 & & \hspace{1cm} -\chi(x)[\phi(x), \chi(x)] +\frac14[\phi(x), \bar{\phi}(x)]^2 -\frac14\eta(x)[\phi(x), \eta(x)] \nn \\
 & & \hspace{1cm} +\sum_{\mu=0}^1\left(\phi(x)-U_{\mu}(x)\phi(x+\hat{\mu})U_{\mu}(x)^{-1}\right)
\left(\bar{\phi}(x)-U_{\mu}(x)\bar{\phi}(x+\hat{\mu})U_{\mu}(x)^{-1}\right) \nn \\
 & & \hspace{1cm} -i\sum_{\mu=0}^1\psi_{\mu}(x)\left(\eta(x)-U_{\mu}(x)\eta(x+\hat{\mu})U_{\mu}(x)^{-1}\right) \nn \\
 & & \hspace{1cm} \left. -\sum_{\mu=0}^1 \psi_{\mu}(x)\psi_{\mu}(x)\left(\bar{\phi}(x)+U_{\mu}(x)\bar{\phi}(x)U_{\mu}(x)^{-1}\right)
\frac{}{} \right],
\label{S_lat_SYM1'}
\eea
which reduces to the continuum expression (\ref{S_cont_SYM}) in the naive continuum limit. 

Concerning the gauge fields, 
the form of the action is essentially same as that of L\"uscher's lattice action of the four-dimensional chiral ${\rm U}(1)$ 
gauge theory~\cite{luscher}. 
Note that the Boltzmann weight $\exp[-S^{\rm LAT}_{\rm SYM}]$ is smooth and infinitely differentiable with respect to 
the lattice fields, and gives no contribution from the field configurations not admissible. 
In this way, the gauge field configurations are effectively restricted to the admissible ones with the smoothness, 
and the degeneracy of the vacua is resolved to single out 
the vacuum $U_{01}(x) =\id_N$ without spoiling the $Q$-supersymmetry~\cite{sugino2}. 

The SYM part exactly coincides to the pure SYM model discussed in~\cite{sugino2}, after renaming the auxiliary field as
\beq
H(x) \equiv iD(x) +\frac{i}{2}\widehat{\Phi}(x). 
\eeq
 
For the case $G={\rm U}(N)$, the FI term can be introduced to the action as the $Q$-exact form 
\beq
Q \kappa \sum_x \tr \left(-i\chi(x)\right) 
  = \kappa \sum_x \tr \left[ D(x) + \frac12 \whPhi(x)\right] 
\eeq
with the counterpart of the imaginary $\vartheta$-term ($\vartheta =2\pi i\kappa$) accompanied. 
Similarly to the continuum case, to compensate the imaginary $\vartheta$-term, 
we can independently add the topological quantity on the lattice 
\beq
-\frac{\vartheta-2\pi i\kappa}{2\pi} \sum_x \tr \, \ln U_{01}(x), 
\eeq
which is $Q$-invariant due to the topological property. 
In fact, for the case of finite periodic lattices or the infinite lattice, 
\bea
Q \sum_x \tr \, \ln U_{01}(x) & = & \sum_x \tr \left[U_{01}(x)^{-1} Q U_{01}(x)\right] \nn \\
 & = & i \sum_x \tr \left[\psi_0(x) +\psi_1(x+\hat{0}) -\psi_0(x+\hat{1}) -\psi_1(x)\right] = 0. 
\eea 
Note that the logarithm of the plaquette variables is guaranteed to be well-defined 
by imposing the admissibility condition stronger than (\ref{admissibility_U(N)}) with 
\beq
0< \epsilon <1. 
\label{admissibility_theta}
\eeq 
Combining these, we can incorporate the $Q$-invariant FI and $\vartheta$-terms into the lattice action as 
\beq
S^{\rm LAT}_{{\rm FI}, \, \vartheta} = Q \kappa \sum_x \tr \left(-i\chi(x)\right) 
-\frac{\vartheta-2\pi i\kappa}{2\pi} \sum_x \tr \, \ln U_{01}(x) 
\label{FI_theta_lat}
\eeq
by choosing $\epsilon$ as (\ref{admissibility_theta}). 

Although the action (\ref{S_lat_SYM1'}) has noncompact flat directions with respect to $\phi(x), \bar{\phi}(x)$, 
it can be lifted by introducing suitable couplings to the matter sector as discussed in~\cite{hanany-hori,hori-tong}.  
It will also be possible to construct the lattice SYM part with the compact Higgs fields, similarly to Ref.~\cite{sugino4}. 
(See also \cite{suzuki-taniguchi}.)  

\subsection{Matter Part of Lattice Theory}
In order to latticize the matter part, we introduce the covariant forward (backward) difference operators $D_\mu$ ($D_\mu^*$) by 
\bea
aD_\mu\Phi_{+I}(x) & = & U_{\mu}(x)\Phi_{+I}(x+\hat{\mu}) -\Phi_{+I}(x), \nn \\
aD_\mu^*\Phi_{+I}(x) & = & \Phi_{+I}(x) - U_{\mu}(x-\hat{\mu})^{-1}\Phi_{+I}(x-\hat{\mu}), \nn \\
aD_\mu\Phi_{+I}(x)^\dagger & = & \Phi_{+I}(x+\hat{\mu})^\dagger U_{\mu}(x)^{-1} -\Phi_{+I}(x)^\dagger, \nn \\aD_\mu^*\Phi_{+I}(x)^\dagger & = & \Phi_{+I}(x)^\dagger -\Phi_{+I}(x-\hat{\mu})^\dagger U_{\mu}(x-\hat{\mu}), \\
aD_\mu \Phi_{-I'}(x) & = & \Phi_{-I'}(x+\hat{\mu}) U_{\mu}(x)^{-1} -\Phi_{-I'}(x), \nn \\
aD_\mu^*\Phi_{-I'}(x) & = & \Phi_{-I'}(x)-\Phi_{-I'}(x-\hat{\mu})U_{\mu}(x-\hat{\mu}), \nn \\
aD_\mu\Phi_{-I'}(x)^\dagger & = &U_{\mu}(x)\Phi_{-I'}(x+\hat{\mu})^\dagger -\Phi_{-I'}(x)^\dagger, \nn \\
aD_\mu^*\Phi_{-I'}(x)^\dagger & = &\Phi_{-I'}(x)^\dagger -U_{\mu}(x-\hat{\mu})^{-1}\Phi_{-I'}(x-\hat{\mu})^\dagger, 
\eea
and denote their symmetric and anti-symmetric combinations as 
\beq
D_\mu^S \equiv \frac12 (D_\mu + D_\mu^*), \qquad D_\mu^A \equiv \frac12 (D_\mu -D_\mu^*). 
\eeq
For the case $n_+=n_-(\equiv n)$, 
the $Q$-supersymmetry transformations for the matter fields (\ref{Qsusy_cont_mat+}), (\ref{Qsusy_cont_mat-}) can be 
realized on the lattice as follows: 
\bea
 & & Q\phi_{+I}(x) = -\psi_{+IL}(x), \qquad Q\psi_{+IL}(x)=-\left (\phi(x) -\twm_{+I}\right)\phi_{+I}(x), \nn \\
 & & Q\psi_{+IR}(x) = a(D_0^S +iD_1^S)\phi_{+I}(x) +\sum_{\mu=0}^1 raD_\mu^A\phi_{-I}(x)^\dagger +F_{+I}(x), \nn \\
 & & QF_{+I}(x) = \left(\phi(x)-\twm_{+I}\right)\psi_{+IR}(x) +a(D_0^S +iD_1^S)\psi_{+IL}(x) 
+\sum_{\mu=0}^1 raD_\mu^A\bar{\psi}_{-IR}(x) \nn \\
 & & \hspace{1cm}-i\frac12\left(\psi_0(x)U_0(x)\phi_{+I}(x+\hat{0}) +U_0(x-\hat{0})^{-1}\psi_0(x-\hat{0})\phi_{+I}(x-\hat{0})\right) \nn \\
 & &  \hspace{1cm}+\frac12\left(\psi_1(x)U_1(x)\phi_{+I}(x+\hat{1}) +U_1(x-\hat{1})^{-1}\psi_1(x-\hat{1})\phi_{+I}(x-\hat{1})\right) \nn \\
 & & \hspace{1cm}-\sum_{\mu=0}^1 \frac{ir}{2}\left(\psi_{\mu}(x)U_{\mu}(x)\phi_{-I}(x+\hat{\mu})^\dagger 
-U_{\mu}(x-\hat{\mu})^{-1}\psi_{\mu}(x-\hat{\mu})\phi_{-I}(x-\hat{\mu})^\dagger\right), \nn \\
 & & Q\phi_{+I}(x)^\dagger = -\bar{\psi}_{+IR}(x), \qquad Q\bar{\psi}_{+IR}(x) = \phi_{+I}(x)^\dagger \left(\phi(x)-\twm_{+I}\right), \nn \\
 & & Q\bar{\psi}_{+IL}(x) = a(D_0^S-iD_1^S)\phi_{+I}(x)^\dagger +\sum_{\mu=0}^1 ra D_{\mu}^A\phi_{-I}(x) +F_{+I}(x)^\dagger, \nn \\
 & & QF_{+I}(x)^\dagger = -\bar{\psi}_{+IL}(x)\left(\phi(x)-\twm_{+I}\right) +a(D_0^S-iD_1^S)\bar{\psi}_{+IR}(x) 
+\sum_{\mu=0}^1 raD_\mu^A\psi_{-IL}(x) \nn \\
 & & \hspace{1cm}+i\frac12\left(\phi_{+I}(x+\hat{0})^\dagger U_0(x)^{-1}\psi_0(x) 
+\phi_{+I}(x-\hat{0})^\dagger\psi_0(x-\hat{0})U_0(x-\hat{0})\right) \nn \\
 & & \hspace{1cm} +\frac12\left(\phi_{+I}(x+\hat{1})^\dagger U_1(x)^{-1}\psi_1(x) 
+\phi_{+I}(x-\hat{1})^\dagger\psi_1(x-\hat{1})U_1(x-\hat{1})\right) \nn \\
 & & \hspace{1cm}+\sum_{\mu=0}^1 \frac{ir}{2}\left(\phi_{-I}(x+\hat{\mu})U_{\mu}(x)^{-1}\psi_{\mu}(x) 
 -\phi_{-I}(x-\hat{\mu})\psi_{\mu}(x-\hat{\mu})U_{\mu}(x-\hat{\mu})\right), 
\eea
\bea
 & & Q\phi_{-I}(x) = -\psi_{-IL}(x), \qquad Q\psi_{-IL}(x) = \phi_{-I}(x)\left(\phi(x)-\twm_{-I}\right), \nn \\
 & & Q\psi_{-IR}(x) = a(D_0^S +iD_1^S)\phi_{-I}(x) + \sum_{\mu=0}^1 raD_\mu^A\phi_{+I}(x)^\dagger + F_{-I}(x), \nn \\
 & & QF_{-I}(x) = -\psi_{-IR}(x)\left(\phi(x)-\twm_{-I}\right) +a(D_0^S+iD_1^S)\psi_{-IL}(x) 
+\sum_{\mu=0}^1 ra D_\mu^A\bar{\psi}_{+IR}(x) \nn \\
 & & \hspace{1cm}+i\frac12\left(\phi_{-I}(x+\hat{0}) U_0(x)^{-1}\psi_0(x) +\phi_{-I}(x-\hat{0})\psi_0(x-\hat{0})U_0(x-\hat{0})\right) \nn \\
 & & \hspace{1cm} -\frac12\left(\phi_{-I}(x+\hat{1}) U_1(x)^{-1}\psi_1(x) +\phi_{-I}(x-\hat{1})\psi_1(x-\hat{1})U_1(x-\hat{1})\right) \nn \\
 & & \hspace{1cm}+\sum_{\mu=0}^1 \frac{ir}{2} \left(\phi_{+I}(x+\hat{\mu})^\dagger U_{\mu}(x)^{-1}\psi_{\mu}(x) 
 -\phi_{+I}(x-\hat{\mu})^\dagger\psi_{\mu}(x-\hat{\mu})U_{\mu}(x-\hat{\mu})\right), \nn \\
 & & Q\phi_{-I}(x)^\dagger = -\bar{\psi}_{-IR}(x), \qquad Q\bar{\psi}_{-IR}(x) =-\left(\phi(x)-\twm_{-I}\right)\phi_{-I}(x)^\dagger, \nn \\
 & & Q\bar{\psi}_{-IL}(x) = a(D_0^S-iD_1^S)\phi_{-I}(x)^\dagger +\sum_{\mu=0}^1 ra D_\mu^A\phi_{+I}(x) +F_{-I}(x)^\dagger , \nn \\
 & & QF_{-I}(x)^\dagger = \left(\phi(x)-\twm_{-I}\right)\bar{\psi}_{-IL}(x) +a(D_0^S-iD_1^S)\bar{\psi}_{-IR}(x) 
+\sum_{\mu=0}^1 raD_\mu^A\psi_{+IL}(x) \nn \\
 & & \hspace{1cm}-i\frac12\left(\psi_0(x)U_0(x)\phi_{-I}(x+\hat{0})^\dagger 
+U_0(x-\hat{0})^{-1}\psi_0(x-\hat{0})\phi_{-I}(x-\hat{0})^\dagger\right) \nn \\
 & & \hspace{1cm} -\frac12\left(\psi_1(x)U_1(x)\phi_{-I}(x+\hat{1})^\dagger 
+U_1(x-\hat{1})^{-1}\psi_1(x-\hat{1})\phi_{-I}(x-\hat{1})^\dagger\right) \nn \\
 & &  \hspace{1cm}-\sum_{\mu=0}^1 \frac{ir}{2}\left(\psi_{\mu}(x)U_{\mu}(x)\phi_{+I}(x+\hat{\mu}) 
-U_{\mu}(x-\hat{\mu})^{-1}\psi_{\mu}(x-\hat{\mu})\phi_{+I}(x-\hat{\mu})\right). 
\eea
Here, $r$ is a real positive parameter, and the Wilson terms containing $r$ are necessary to suppress the 
contribution from the species doublers appearing in both of the fermionic and bosonic degrees of freedom\footnote{
It is consistent with the supersymmetry preserved on the lattice. 
The species doublers can be explicitly seen from the poles of 
the propagators (\ref{prop_lat_pert}) -- (\ref{lat_mom}).}. From the structure of the Wilson terms, 
which connect the fundamentals and anti-fundamentals in each flavor $I$, we must take $n_+=n_-$. 
The nilpotency of $Q$ holds similarly to the continuum case, except for the auxiliary fields: 
\bea
Q^2 F_{+I}(x) & = & \left(\phi(x) -\twm_{+I}\right) F_{+I}(x) 
-\left(\twm_{+I} -\twm_{-I}\right)\sum_{\mu =0}^1 ra D_{\mu}^A \phi_{-I}(x)^\dagger,  \nn \\
Q^2 F_{+I}(x)^\dagger & = & -F_{+I}(x)^\dagger \left(\phi(x) -\twm_{+I}\right) 
+\left(\twm_{+I} -\twm_{-I}\right) \sum_{\mu =0}^1 ra D_{\mu}^A \phi_{-I}(x), \nn \\
Q^2 F_{-I}(x) & = & -F_{-I}(x) \left(\phi(x) -\twm_{-I}\right) 
-\left(\twm_{+I} -\twm_{-I}\right) \sum_{\mu =0}^1 ra D_{\mu}^A \phi_{+I}(x)^\dagger, \nn \\
Q^2 F_{-I}(x)^\dagger & = & \left(\phi(x) -\twm_{-I}\right) F_{-I}(x)^\dagger 
+\left(\twm_{+I} -\twm_{-I}\right)\sum_{\mu =0}^1 ra D_{\mu}^A \phi_{+I}(x), 
\eea
where the contribution from the Wilson terms violates the nilpotency. 
If we focus on the case 
\beq
\twm_{+I} =\twm_{-I} (\equiv \twm_I), 
\label{twm_lat}
\eeq
the violation disappears and the $Q$-supersymmetry becomes entirely 
nilpotent up to the combination of the infinitesimal gauge transformation with 
the parameter $\phi(x)$ and the infinitesimal flavor rotations with the parameters $\twm_I$ acting as 
\beq
\delta\Phi_{\pm I} = \mp \twm_{I}\Phi_{\pm I}, \qquad \delta\Phi_{\pm I}^\dagger = \pm \twm_{I}\Phi_{\pm I}^\dagger. 
\eeq

Then, the matter parts of the action (\ref{S_cont_mat+}), (\ref{S_cont_mat-}) can be transcribed on the lattice as the $Q$-exact form: 
\bea
S^{\rm LAT}_{{\rm mat}, +\twm} & = & 
Q \sum_x \sum_{I=1}^n\frac12 \left[\bar{\psi}_{+IL}(x)\left\{a(D_0^S+iD_1^S)\phi_{+I}(x) 
+\sum_{\mu=0}^1 raD_{\mu}^A\phi_{-I}(x)^\dagger -F_{+I}(x)\right\} \right. \nn \\
 & & \hspace{2cm} +\left\{a(D_0^S-iD_1^S)\phi_{+I}(x)^\dagger +\sum_{\mu=0}^1 raD_{\mu}^A\phi_{-I}(x) -F_{+I}(x)^\dagger\right\} 
\psi_{+IR}(x) \nn \\
 & & \hspace{2cm} +\bar{\psi}_{+IR}(x)\left(\bar{\phi}(x)-\twm_{+I}^*\right)\phi_{+I}(x)  
-\phi_{+I}(x)^\dagger \left(\bar{\phi}(x)-\twm_{+I}^*\right)\psi_{+IL}(x) \nn \\
 & & \hspace{2cm} \left. +2i\phi_{+I}(x)^\dagger \chi(x)\phi_{+I}(x) \frac{}{} \right], 
\label{S_lat_mat+}\\
S^{\rm LAT}_{{\rm mat}, -\twm} & = & Q \sum_x \sum_{I=1}^n \frac12 \left[\left\{a(D_0^S+iD_1^S)\phi_{-I}(x) 
+\sum_{\mu=0}^1 raD_{\mu}^A\phi_{+I}(x)^\dagger -F_{-I}(x)\right\} \bar{\psi}_{-IL}(x) \right. \nn \\
 & & \hspace{2cm} +\psi_{-IR}(x)\left\{a(D_0^S -iD_1^S)\phi_{-I}(x)^\dagger 
+\sum_{\mu=0}^1 ra D_{\mu}^A\phi_{+I}(x) -F_{-I}(x)^\dagger\right\} \nn \\
 & & \hspace{2cm} -\phi_{-I}(x)\left(\bar{\phi}(x)-\twm_{-I}^*\right)\bar{\psi}_{-IR}(x) 
+\psi_{-IL}(x)\left(\bar{\phi}(x)-\twm_{-I}^*\right)\phi_{-I}(x)^\dagger  \nn \\
 & & \hspace{2cm} \left. -2i\phi_{-I}(x)\chi(x)\phi_{-I}(x)^\dagger \frac{}{} \right].
\label{S_lat_mat-} 
\eea
Due to the Wilson terms, the flavor symmetry of the lattice actions (\ref{S_lat_mat+}), (\ref{S_lat_mat-}) reduces to 
${\rm U}(1)^n$, the diagonal subgroup of ${\rm U}(1)^n\times {\rm U}(1)^n$ of the continuum case for $n_+=n_-\equiv n$. 
Thus, the actions are guaranteed to be $Q$-invariant, when the flavor rotation generated by $Q^2$ falls into the diagonal 
${\rm U}(1)^n$. Again, this is nothing but the case of (\ref{twm_lat}). 
In what follows, we consider the case~(\ref{twm_lat}).  
Note that we can still freely choose the anti-holomorphic twisted masses $\twm_{+I}^*, \twm_{-I}^*$. 

Also, for the superpotential terms (\ref{S_cont_pot}), it is straightforward to write down the lattice counterpart 
\bea
S^{\rm LAT}_{\rm pot} & = & Q \sum_x \sum_{i=1}^N\sum_{I=1}^n\left[-\frac{\der W(\phi_+, \phi_-)}{\der\phi_{+Ii}(x)}\psi_{+IRi}(x) 
-\bar{\psi}_{+ILi}(x)\frac{\der \bar{W}(\phi_+^\dagger, \phi_-^\dagger)}{\der \phi_{+Ii}(x)^*} \right. \nn \\
 & & \hspace{2.5cm} \left. -\psi_{-IRi}(x)\frac{\der W(\phi_+, \phi_-)}{\der \phi_{-Ii}(x)} 
-\frac{\der \bar{W}(\phi_+^\dagger, \phi_-^\dagger)}{\der \phi_{-Ii}(x)^*}\bar{\psi}_{-ILi}(x) \right]. 
\label{S_lat_pot} 
 \eea
Note that all the terms are not exactly holomorphic or anti-holomorphic, because the holomorphic and anti-holomorphic fields 
are mixed at the order $\cO(a)$ through the contribution from the Wilson terms.

\setcounter{equation}{0}
\section{${\rm U}(1)_A$ Anomaly}
\label{sec:anomaly}
We analyze the anomaly for the ${\rm U}(1)_A$ R-symmetry in the system with the twisted mass terms introduced, 
for both cases of the continuum and the lattice, i.e. 
\bea
S^{(E)}_{{\rm 2DSQCD}, \twm} & \equiv & S^{(E)}_{\rm 2DSYM} + S^{(E)}_{{\rm mat}, +\twm} + S^{(E)}_{{\rm mat}, -\twm}  \qquad 
\mbox{for the continuum case},  \\
S^{\rm LAT}_{{\rm 2DSQCD}, \twm} & \equiv & S^{\rm LAT}_{\rm 2DSYM} + S^{\rm LAT}_{{\rm mat}, +\twm} 
+ S^{\rm LAT}_{{\rm mat}, -\twm}  \qquad \mbox{for the lattice case}. 
\label{S_lat_SQCD_twm}
\eea
In this section, we consider the case $G={\rm U}(N)$. 

In particular, although the $Q$-invariant lattice action presented in the previous section 
is defined in the case $n_+=n_-$, by sending some of the anti-holomorphic twisted masses ($\twm_{+I}^*$'s or $\twm_{-I}^*$'s) 
to the infinity, 
we show that the anomaly for the case $n_+\neq n_-$ is correctly obtained from our lattice action. 
  
\subsection{${\rm U}(1)_A$ Anomaly in the Continuum Theory}
Without taking into account the quantum effect, the Ward-Takahashi (WT) identity for the ${\rm U}(1)_A$ rotation is naively 
derived as
\beq
\vev{\der_{\mu}j_{A\mu}(x)} \naiveeq \vev{M(x)},
\label{anomaly_naive}
\eeq
where $j_{A\mu}(x)$ is the corresponding Noether current, 
and $M(x)$ represents the explicit breaking by the twisted mass terms:  
\bea
 j_{A\mu}(x)  & = &  j_{A\mu}^{{\rm SYM}}(x) 
+ j_{A\mu}^{{\rm mat}}(x),  \\
 j_{A\mu}^{{\rm SYM}} & \equiv & \frac{1}{g^2}\tr \left[2\phi\cD_{\mu}\bar{\phi} -2(\cD_\mu\phi)\bar{\phi} 
 +i\eta\psi_{\mu} +2i\epsilon_{\mu\nu}\chi\psi_{\nu}\right] \qquad (\epsilon_{01}=-\epsilon_{10}=+1), \nn \\
j_{A0}^{{\rm mat}} &\equiv  & \sum_{I=1}^{n_+} \left(\bar{\psi}_{+IL} \, \psi_{+IL}-\bar{\psi}_{+IR} \, \psi_{+IR}\right) 
 +\sum_{I'=1}^{n_-}\left(-\psi_{-I'L} \, \bar{\psi}_{-I'L} +\psi_{-I'R} \, \bar{\psi}_{-I'R}\right), \nn \\
j_{A1}^{{\rm mat}} & \equiv  & \sum_{I=1}^{n_+} \left(i\bar{\psi}_{+IL} \, \psi_{+IL} +i\bar{\psi}_{+IR} \, \psi_{+IR}\right) 
 +\sum_{I'=1}^{n_-}\left(-i\psi_{-I'L} \, \bar{\psi}_{-I'L} -i\psi_{-I'R} \, \bar{\psi}_{-I'R}\right), \nn \\
 M(x)  & = & M_B(x) + M_F(x),  \\
 M_B &\equiv & 2\sum_{I=1}^{n_+} \left( \twm_{+I} \, \phi_{+I}^\dagger \, \bar{\phi} \, \phi_{+I}
-\twm_{+I}^* \, \phi_{+I}^\dagger \, \phi \, \phi_{+I} \right) \nn \\
 & & +2\sum_{I'=1}^{n_-} \left( \twm_{-I'} \, \phi_{-I'} \, \bar{\phi} \, \phi_{-I'}^\dagger 
-\twm_{-I'}^* \, \phi_{-I'} \, \phi \, \phi_{-I'}^\dagger \right), \nn \\
 M_F & \equiv & 2\sum_{I=1}^{n_+} \left( \twm_{+I} \, \bar{\psi}_{+IL} \, \psi_{+IR} -\twm_{+I}^* \, \bar{\psi}_{+IR} \, \psi_{+IL} \right) \nn \\
 & & +2\sum_{I'=1}^{n_-} \left( \twm_{-I'} \, \psi_{-I'R} \, \bar{\psi}_{-I'L} -\twm_{-I'}^* \, \psi_{-I'L} \, \bar{\psi}_{-I'R} \right). \nn
\eea
 
To derive the anomaly potentially arising in (\ref{anomaly_naive}), 
we perform the perturbative computation for the matter multiplets up to the first order.   
Since the gaugino belongs to the adjoint representation of $G$, the SYM part does not contribute to the anomaly. 
First, let us calculate $\dvev{j^{\rm mat}_{A\mu}(x)}$, 
where $\dvev{\cdot}$ means the expectation value with respect to the integration of the matter fields 
under the action $S^{(E)}_{{\rm mat}, +\twm} + S^{(E)}_{{\rm mat}, -\twm}$.  
The propagators are given by 
\bea
 \vev{\phi_{+Ii}(x)\phi_{+Jj}(y)^*}_0 & = & \delta_{IJ}\delta_{ij} \int \frac{\dd^2p}{(2\pi)^2} \, e^{ip\cdot (x-y)} \Delta_{+I}(p), 
\nn \\
\vev{\phi_{-I'i}(x)^*\phi_{-J'j}(y)}_0 & = & \delta_{I'J'}\delta_{ij} \int \frac{\dd^2p}{(2\pi)^2} \, e^{ip\cdot (x-y)} \Delta_{-I'}(p), \nn
\\
\vev{\psi_{+I\alpha i}(x)\bar{\psi}_{+J\beta j}(y)}_0 & = &  \delta_{IJ}\delta_{ij} \int \frac{\dd^2p}{(2\pi)^2} \, e^{ip\cdot (x-y)} 
\left(T_{+I}(p)\right)_{\alpha\beta}, \nn \\
\vev{\bar{\psi}_{-I'\alpha i}(x)\psi_{-J'\beta j}(y)}_0 & = & \delta_{IJ}\delta_{ij} \int \frac{\dd^2p}{(2\pi)^2} \, e^{ip\cdot (x-y)} 
\left(T_{-I'}(p)\right)_{\alpha\beta},  
\eea
with $\alpha, \beta$ running over the indices $L, R$, and 
\bea
\Delta_{+I}(p) \equiv \frac{1}{p^2+\twm_{+I}\twm_{+I}^*}, & & 
T_{+I}(p) \equiv -\Delta_{+I}(p) \left(\begin{array}{cc} ip_0+p_1 & \twm_{+I} \\ \twm_{+I}^* & ip_0-p_1\end{array}\right), \nn \\
\Delta_{-I'}(p) \equiv \frac{1}{p^2+\twm_{-I'}\twm_{-I'}^*}, & & 
T_{-I'}(p) \equiv -\Delta_{-I'}(p) \left(\begin{array}{cc} ip_0+p_1 & \twm_{-I'}^* \\ \twm_{-I'} & ip_0-p_1\end{array}\right).
\eea
In the zeroth order of the perturbation, it is easy to see that $\dvev{j^{\rm mat}_{A\mu}(x)}$ vanishes. 
Up to the first order, we have the result 
\bea
\dvev{j^{\rm mat}_{A\mu}(x)} & = & \int \frac{\dd^2k}{(2\pi)^2}\, e^{ik\cdot x} \, \epsilon_{\mu\rho}
\left(\sum_{I=1}^{n_+}\Pi_{+I \rho\nu}(k) -\sum_{I'=1}^{n_-}\Pi_{-I' \rho\nu}(k)\right) \tr \, \wt{A}_{\nu}(k) \nn \\
 & & \hspace{-0.2cm} +\sum_{I=1}^{n_+}\frac{1}{4\pi}\int \frac{\dd^2k}{(2\pi)^2}\, e^{ik\cdot x} \int_0^1\dd\alpha \, 
\frac{1}{\alpha(1-\alpha)k^2+\twm_{+I}\twm_{+I}^*} \nn \\
 & & \hspace{0.3cm} \times\tr\left[\twm_{+I}^*\left(ik_{\mu}+(1-2\alpha)\epsilon_{\mu\nu}k_\nu\right)\wt{\phi}(k) 
 +\twm_{+I}\left(-ik_\mu +(1-2\alpha)\epsilon_{\mu\nu}k_\nu\right)\wt{\bar{\phi}}(k)\right] \nn \\
 & & \hspace{-0.2cm}-\sum_{I'=1}^{n_-}\frac{1}{4\pi}\int \frac{\dd^2k}{(2\pi)^2}\, e^{ik\cdot x} \int_0^1\dd\alpha \, 
\frac{1}{\alpha(1-\alpha)k^2+\twm_{-I'}\twm_{-I'}^*} \nn \\
 & & \hspace{0.3cm} \times\tr\left[\twm_{-I'}^*\left(-ik_{\mu}+(1-2\alpha)\epsilon_{\mu\nu}k_\nu\right)\wt{\phi}(k) 
 +\twm_{-I'}\left(ik_\mu +(1-2\alpha)\epsilon_{\mu\nu}k_\nu\right)\wt{\bar{\phi}}(k)\right]. \nn \\
 & & \label{jmat_cont}
\eea
Here, the fields of the SYM sector are expressed as their Fourier modes, and the Feynman parameter $\alpha$ is introduced. 
Also, 
\bea
\Pi_{+I\rho\nu}(k) & \equiv & -\frac{1}{\pi}\int_0^1\dd\alpha \, \frac{\alpha(1-\alpha)}{\alpha(1-\alpha)k^2 +\twm_{+I}\twm_{+I}^*}
\left(\delta_{\rho\nu}k^2-k_{\rho}k_{\nu}\right) + \frac{1}{2\pi}\delta_{\rho\nu}, \nn \\
 \Pi_{-I'\rho\nu}(k) & \equiv & -\frac{1}{\pi}\int_0^1\dd\alpha \, \frac{\alpha(1-\alpha)}{\alpha(1-\alpha)k^2 +\twm_{-I'}\twm_{-I'}^*}
\left(\delta_{\rho\nu}k^2-k_{\rho}k_{\nu}\right) + \frac{1}{2\pi}\delta_{\rho\nu}. 
\label{vp_cont}
\eea
$\Pi_{+I\rho\nu}(k)$ ($\Pi_{-I'\rho\nu}(k)$) is the vacuum polarization tensor for the overall ${\rm U}(1)$ gauge field, 
which comes from the loop of the (anti-)fundamental fermions. 
The gauge invariance requires 
\bea
k_{\nu}\Pi_{+I\rho\nu}(k) = k_{\nu}\Pi_{-I'\rho\nu}(k)=0. 
\label{gaugeinv}
\eea
However, the expression of (\ref{vp_cont}) does not satisfy it. 
To meet (\ref{gaugeinv}), we remove the last terms in the r.h.s. of (\ref{vp_cont}) and redefine by 
\bea
\Pi_{+I\rho\nu}^{\rm (new)}(k) & \equiv & \Pi_{+I\rho\nu}(k) -\frac{1}{2\pi}\delta_{\rho\nu}, \label{vp+_new} \\
\Pi_{-I'\rho\nu}^{\rm (new)}(k) & \equiv & \Pi_{-I'\rho\nu}(k) -\frac{1}{2\pi}\delta_{\rho\nu}. \label{vp-_new}
\eea
(\ref{vp+_new}) corresponds to the procedure to add an appropriate local counter term 
to the effective action, which is obtained after integrating out 
the fundamental matters in $S^{(E)}_{{\rm mat}, +\twm}$, to recover the gauge invariance. 
Also, (\ref{vp-_new}) corresponds to the modification to the effective action after the integration of the anti-fundamental matters 
in $S^{(E)}_{{\rm mat},-\twm}$. 
In fact, denoting the overall ${\rm U}(1)$ gauge currents of $G={\rm U}(N)$ derived from $S^{(E)}_{{\rm mat}, \pm \twm}$ as 
$j^{{\rm mat}, \pm}_{\mu}(x)$, the ${\rm U}(1)_A$ current can be expressed as 
\beq
j^{\rm mat}_{A\mu} = -\epsilon_{\mu\nu}j^{{\rm mat},+}_\nu + \epsilon_{\mu\nu}j^{{\rm mat},-}_\nu
\eeq
with the bosonic terms neglected in the r.h.s. 
Since $\dvev{j^{{\rm mat}, \pm}_{\mu}(x)}$ are obtained by differentiating the matter effective actions 
with respect to the overall ${\rm U}(1)$ gauge field $A^{\rm U(1)}$: 
\bea
\dvev{j^{{\rm mat}, \pm}_{\mu}(x)} & = & \frac{\delta W_{\pm}[\text{SYM fields}]}{\delta A^{\rm U(1)}_{\mu}(x)}, \\
e^{-W_+[\text{SYM fields}]} & \equiv &  
\int \left(\prod_{I=1}^{n_+} \cD\Phi_{+I}\cD\Phi_{+I}^\dagger\right) \, e^{-S^{(E)}_{{\rm mat}, +\twm}}, \nn \\
e^{-W_-[\text{SYM fields}]} & \equiv  & 
\int \left( \prod_{I'=1}^{n_-} \cD\Phi_{-I'}\cD\Phi_{-I'}^\dagger\right) \, e^{-S^{(E)}_{{\rm mat}, -\twm}}, \nn 
\eea
the modification (\ref{vp+_new}), (\ref{vp-_new}) is equivalent to adding the local counter terms to the effective actions as 
\beq
 W_{\pm}^{\rm (new)}[\text{SYM fields}] \equiv W_{\pm}[\text{SYM fields}] + \frac{n_\pm}{4\pi} \int \dd^2x \,
\tr \left(A_{\mu}(x)^2\right).
\eeq 
For $\dvev{j^{\rm mat}_{A\mu}(x)}$ modified by replacing $\Pi_{+I\rho\nu}(k)$ and $\Pi_{-I'\rho\nu}(k)$ in (\ref{jmat_cont}) with 
$\Pi^{\rm (new)}_{+I\rho\nu}(k)$ and $\Pi^{\rm (new)}_{-I'\rho\nu}(k)$, we have 
\bea
\dvev{\der_{\mu}j^{\rm mat}_{A\mu}(x)} & = & -\frac{1}{\pi}(n_+ -n_-) \, \tr \, F_{01}(x) 
  + \int\frac{\dd^2k}{(2\pi)^2}\, e^{ik\cdot x} \int_0^1\dd\alpha \left\{\frac{1}{\pi}\tr \, \wt{F}_{01}(k)  \right. \nn \\
& & \hspace{1cm}\times \left[\sum_{I=1}^{n_+}\frac{\twm_{+I}\twm_{+I}^*}{\alpha(1-\alpha)k^2+\twm_{+I}\twm_{+I}^*} 
-\sum_{I'=1}^{n_-}\frac{\twm_{-I'}\twm_{-I'}^*}{\alpha(1-\alpha)k^2+\twm_{-I'}\twm_{-I'}^*}\right]  \nn \\
  & &  \hspace{0.7cm}+\sum_{I=1}^{n_+} \frac{1}{4\pi}\frac{k^2}{\alpha(1-\alpha)k^2+\twm_{+I}\twm_{+I}^*} \,
  \tr \left[-\twm_{+I}^*\wt{\phi}(k) +\twm_{+I}\wt{\bar{\phi}}(k)\right] \nn \\
  & & \hspace{0.7cm}\left. +\sum_{I=1'}^{n_-} \frac{1}{4\pi}\frac{k^2}{\alpha(1-\alpha)k^2+\twm_{-I'}\twm_{-I'}^*} \,
  \tr \left[-\twm_{-I'}^*\wt{\phi}(k) +\twm_{-I'}\wt{\bar{\phi}}(k)\right] \right\}. \nn \\
  & & 
\label{anomaly_cont_lhs}  
\eea
  
Next, let us compute $\dvev{M(x)}$.   
It is easily seen that the zeroth order contribution of $\dvev{M_F(x)}$
vanishes. Since $M_B(x)$ already contains the SYM fields $\phi(x),
\bar{\phi}(x)$, 
the contributions from the first order of $\dvev{M_F(x)}$ and from the zeroth order of
$\dvev{M_B(x)}$ are to be compared with (\ref{anomaly_cont_lhs}). 
The contribution turns out to be equal to the second term in (\ref{anomaly_cont_lhs}). Thus, we obtain 
\beq
\dvev{\der_{\mu}j^{\rm mat}_{A\mu}(x)}  = -\frac{1}{\pi}(n_+ -n_-) \, \tr\, F_{01}(x) +  \dvev{M(x)}, 
\label{anomaly_cont'}
\eeq
which leads to the anomalous WT identity
\beq
\vev{\der_{\mu}j_{A\mu}(x)}=-\frac{1}{\pi}(n_+ -n_-)\vev{\tr \, F_{01}(x)} +  \vev{M(x)}
\label{anomaly_cont}
\eeq
instead of (\ref{anomaly_naive}).

\subsection{${\rm U}(1)_A$ Anomaly in the Lattice Theory}
Since the variables in the lattice actions discussed in section~\ref{sec:lat_2DSQCD} 
are dimensionless, for the perturbative calculation of the anomaly, it is convenient to rescale them to 
assign the dimensions same as those in the continuum theory: 
\bea
 & & \phi(x) \to a \, \phi(x), \qquad \bar{\phi}(x) \to a \, \bar{\phi}(x), \nn \\
 & & \psi_{\mu}(x) \to a^{3/2} \psi_{\mu}(x), \qquad \chi(x) \to a^{3/2}\chi(x), \qquad \eta(x) \to a^{3/2} \eta(x), \nn \\
 & & \psi_{\pm I L}(x) \to a^{1/2} \psi_{\pm I L}(x), \qquad \bar{\psi}_{\pm I L}(x) \to a^{1/2} \bar{\psi}_{\pm I L}(x), \nn \\
 & & \psi_{\pm I R}(x) \to a^{1/2} \psi_{\pm I R}(x), \qquad \bar{\psi}_{\pm I R}(x) \to a^{1/2} \bar{\psi}_{\pm I R}(x), \nn \\
 & & F_{\pm I}(x) \to a \, F_{\pm I}(x), \qquad F_{\pm I}(x)^\dagger \to a \, F_{\pm I}(x)^\dagger, 
\eea
also for the twisted masses
\beq
\twm_{I} \to a \, \twm_{I}, \qquad \twm_{\pm I}^* \to a \, \twm_{\pm I}^*. 
\eeq
 
The ${\rm U}(1)_A$-Noether current obtained from the lattice action is 
\beq 
 J_{A\mu}(x)   =   J_{A\mu}^{\rm SYM}(x) + J_{A\mu}^{\rm mat}(x).  
\eeq
Although the SYM part does not contribute the anomaly and not appear in the following calculation, 
its explicit form is presented in appendix~\ref{app:lat_pert} for the completeness. 
We write the contribution from the matter part dividing into the two parts, the $r$-independent part 
$\hat{J}_{A\mu}^{\rm mat}(x)$ and the $r$-dependent part $\check{J}_{A\mu}^{\rm mat}(x)$:
\bea
 J_{A\mu}^{\rm mat}(x) & = & \hat{J}_{A\mu}^{\rm mat}(x)+ \check{J}_{A\mu}^{\rm mat}(x), 
\\
 \hat{J}_{A0}^{\rm mat}(x) & \equiv & \frac12 \sum_{I=1}^n \left[
\bar{\psi}_{+IL}(x) U_0(x)\psi_{+IL}(x+\hat{0}) + \bar{\psi}_{+IL}(x+\hat{0})U_0(x)^{-1}\psi_{+IL}(x) \right. \nn \\
 & & \hspace{1cm} -\bar{\psi}_{+IR}(x+\hat{0})U_0(x)^{-1}\psi_{+IR}(x) -\bar{\psi}_{+IR}(x)U_0(x)\psi_{+IR}(x+\hat{0}) \nn \\
 & & \hspace{1cm} -\psi_{-IL}(x+\hat{0})U_0(x)^{-1}\bar{\psi}_{-IL}(x) -\psi_{-IL}(x)U_0(x)\bar{\psi}_{-IL}(x+\hat{0}) \nn \\
 & & \hspace{1cm} + \psi_{-IR}(x)U_0(x)\bar{\psi}_{-IR}(x+\hat{0}) + \psi_{-IR}(x+\hat{0}) U_0(x)^{-1}\bar{\psi}_{-IR}(x) \nn \\
 & & \hspace{0.5cm} + ia \bar{\psi}_{+IL}(x+\hat{0}) U_0(x)^{-1}\psi_0(x)\phi_{+I}(x) 
+ ia \phi_{+I}(x)^\dagger \psi_0(x)U_0(x)\psi_{+IR}(x+\hat{0}) \nn \\
 & & \hspace{0.5cm} \left. +ia \phi_{-I}(x)\psi_0(x)U_0(x)\bar{\psi}_{-IL}(x+\hat{0}) 
+ia \psi_{-IR}(x+\hat{0}) U_0(x)^{-1}\psi_0(x)\phi_{-I}(x)^\dagger\right], 
\nn \\
\hat{J}_{A1}^{\rm mat}(x) & \equiv &  \frac12 \sum_{I=1}^n \left[
i\bar{\psi}_{+IL}(x)U_1(x)\psi_{+IL}(x+\hat{1}) +i\bar{\psi}_{+IL}(x+\hat{1}) U_1(x)^{-1} \psi_{+IL}(x) \right. \nn \\
 & & \hspace{1cm} +i\bar{\psi}_{+IR}(x+\hat{1}) U_1(x)^{-1}\psi_{+IR}(x) +i\bar{\psi}_{+IR}(x) U_1(x)\psi_{+IR}(x+\hat{1}) \nn \\
 & & \hspace{1cm} -i\psi_{-IL}(x)U_1(x)\bar{\psi}_{-IL}(x+\hat{1}) -i\psi_{-IL}(x+\hat{1})U_1(x)^{-1}\bar{\psi}_{-IL}(x) \nn \\
 & & \hspace{1cm} -i\psi_{-IR}(x) U_1(x)\bar{\psi}_{-IR}(x+\hat{1}) -i\psi_{-IR}(x+\hat{1})U_1(x)^{-1}\bar{\psi}_{-IR}(x) \nn \\
 & & \hspace{0.5cm} -a\bar{\psi}_{+IL}(x+\hat{1}) U_1(x)^{-1}\psi_1(x)\phi_{+I}(x) 
+a\phi_{+I}(x)^\dagger\psi_1(x)U_1(x)\psi_{+IR}(x+\hat{1}) \nn \\
 & & \hspace{0.5cm} \left. -a\phi_{-I}(x)\psi_1(x) U_1(x)\bar{\psi}_{-IL}(x+\hat{1}) 
 +a\psi_{-IR}(x+\hat{1}) U_1(x)^{-1}\psi_1(x)\phi_{-I}(x)^\dagger\right], 
\nn \\
\check{J}_{A\mu}^{\rm mat}(x) & \equiv & \frac{r}{2}\sum_{I=1}^n\left[ 
\bar{\psi}_{+IL}(x)U_\mu(x)\bar{\psi}_{-IR}(x+\hat{\mu}) -\bar{\psi}_{+IL}(x+\hat{\mu}) U_\mu(x)^{-1}\bar{\psi}_{-IR}(x) \right. \nn \\
 & & \hspace{1cm} +\psi_{-IL}(x)U_\mu(x)\psi_{+IR}(x+\hat{\mu}) -\psi_{-IL}(x+\hat{\mu}) U_\mu(x)^{-1}\psi_{+IR}(x) \nn \\
 & & \hspace{1cm} +\bar{\psi}_{+IR}(x)U_\mu(x)\bar{\psi}_{-IL}(x+\hat{\mu}) 
-\bar{\psi}_{+IR}(x+\hat{\mu}) U_\mu(x)^{-1}\bar{\psi}_{-IL}(x) \nn \\
 & & \hspace{1cm} +\psi_{-IR}(x) U_\mu(x) \psi_{+IL}(x+\hat{\mu}) -\psi_{-IR}(x+\hat{\mu}) U_\mu(x)^{-1} \psi_{+IL}(x) \nn \\
 & & \hspace{0.5cm} -ia\bar{\psi}_{+IL}(x+\hat{\mu}) U_\mu(x)^{-1}\psi_\mu(x)\phi_{-I}(x)^\dagger 
-ia\phi_{-I}(x)\psi_\mu(x) U_\mu(x)\psi_{+IR}(x+\hat{\mu}) \nn \\
 & & \hspace{0.5cm} \left. -ia\phi_{+I}(x)^\dagger\psi_\mu(x)U_\mu(x)\bar{\psi}_{-IL}(x+\hat{\mu}) 
-ia\psi_{-IR}(x+\hat{\mu}) U_\mu(x)^{-1}\psi_\mu(x)\phi_{+I}(x) \right].  
\nn 
\eea
$\check{J}_{A\mu}^{\rm mat}(x)$ represents the contribution from the Wilson terms. 

The ${\rm U}(1)_A$ WT-identity is derived from the lattice theory (\ref{S_lat_SQCD_twm}) as 
\beq
\vev{\sum_{\mu=0}^1\nabla_{\mu}^* \hat{J}_{A\mu}(x)} = -\vev{\sum_{\mu=0}^1\nabla_{\mu}^* \check{J}_{A\mu}(x)} 
+\vev{{\cal M}(x)}, 
\label{anomaly_lat}
\eeq
where $\nabla_{\mu}^*$ is the backward difference operator: $\nabla_{\mu}^* f(x) \equiv\frac{1}{a}\left(f(x)-f(x-\hat{\mu})\right)$, 
and 
\bea
 {\cal M}(x)  & = & {\cal M}_B(x) + {\cal M}_F(x),  \\
 {\cal M}_B(x) &\equiv & 2\sum_{I=1}^{n} \left( \twm_{I} \, \phi_{+I}(x)^\dagger\bar{\phi}(x)\phi_{+I}(x)
-\twm_{+I}^* \, \phi_{+I}(x)^\dagger\phi(x)\phi_{+I}(x) \right. \nn \\
 & & \hspace{1cm}\left. +\twm_{I} \, \phi_{-I}(x)\bar{\phi}(x)\phi_{-I}(x)^\dagger 
-\twm_{-I}^* \, \phi_{-I}(x)\phi(x)\phi_{-I}(x)^\dagger \right), \nn \\
 {\cal M}_F(x) & \equiv & 2\sum_{I=1}^{n} \left( \twm_{I} \, \bar{\psi}_{+IL}(x)\psi_{+IR}(x) 
-\twm_{+I}^* \, \bar{\psi}_{+IR}(x)\psi_{+IL}(x) \right. \nn \\
 & & \hspace{1cm}\left. +\twm_{I} \, \psi_{-IR}(x)\bar{\psi}_{-IL}(x) -\twm_{-I}^* \, \psi_{-IL}(x)\bar{\psi}_{-IR}(x) \right). \nn
\eea
Note that, in contrast to the continuum case~(\ref{anomaly_naive}), the formula (\ref{anomaly_lat}) is exact, because the lattice system gives an unambiguous ultra-violet completion. 
In particular, the path integral measure is explicitly given as 
\bea
(\dd\mu) & = & (\dd\mu_{\rm SYM}) \, (\dd\mu_{\rm mat}), \nn \\
(\dd\mu_{\rm SYM}) & \equiv & \prod_x \left[\prod_{\mu=0}^1 \dd U_{\mu}(x)\right]
\prod_{\bA} \dd\psi^{\bA}_0(x) \, \dd\psi^{\bA}_1(x) \, \dd \chi^{\bA}(x) \, \dd\eta^{\bA}(x) \, \dd\phi^{\bA}(x) \, 
 \dd\bar{\phi}^{\bA}(x) \, \dd D^{\bA}(x), \nn \\
(\dd\mu_{\rm mat}) & = & \prod_{I=1}^n  (\dd\mu_{{\rm mat}, +I}) (\dd\mu_{{\rm mat}, -I}), \nn \\
(\dd\mu_{{\rm mat}, \pm I}) & \equiv & \prod_x \prod_{i=1}^N 
\dd\phi_{\pm Ii}(x) \, \dd\phi_{\pm Ii}(x)^* \, \dd\psi_{\pm I Li}(x) \, \dd\psi_{\pm I Ri}(x) \, 
\dd \bar{\psi}_{\pm I Li}(x) \, \dd\bar{\psi}_{\pm I Ri}(x) \nn \\
 & & \hspace{1cm} \times \dd F_{\pm Ii}(x) \, \dd F_{\pm Ii}(x)^*, 
\eea
where $\dd U_{\mu}(x)$ is the Haar measure of the gauge group $G$, the index $\bA$ labels the generators of $G$, 
and the variables with the index $\bA$ represent the expansion coefficients by the generators of $G$: 
\beq
(\text{field})(x) = \sum_{\bA} (\text{field})^{\bA}(x) \, T^{\bA}, \qquad 
\tr \left(T^{\bA} T^{\bB}\right) = \frac12 \, \delta^{\bA\bB}.
\eeq
Each of $(\dd\mu_{\rm SYM})$, $(\dd\mu_{{\rm mat}, +I})$ and $(\dd\mu_{{\rm mat}, -I})$ is invariant under the ${\rm U}(1)_A$ 
rotation. 

We will integrate out the matter multiplets perturbatively to compute the r.h.s. of (\ref{anomaly_lat}). 
First, we separate the matter action $S^{\rm LAT}_{{\rm mat}, +\twm} + S^{\rm LAT}_{{\rm mat}, -\twm}$ into 
the free Gaussian part $S^{\rm LAT-(2)}_{{\rm mat}, \twm}$ and the interaction part 
$S^{\rm LAT-int}_{{\rm mat}, \twm}$, then expand $U_{\mu}(x) = e^{iaA_{\mu}(x)}$ with respect to $A_{\mu}(x)$. 
The propagators are read off from the Gaussian part as 
\bea
\vev{\phi_{+Ii}(x)\phi_{+Jj}(y)^*}_0 &  = & \delta_{IJ}\delta_{ij} \int_{-\pi /a}^{\pi /a} \frac{\dd^2q}{(2\pi)^2} \, e^{iaq\cdot (x-y)} 
\widehat{\Delta}_{+I}(q), \nn \\
\vev{\phi_{-Ii}(x)^* \phi_{-Jj}(y)}_0 & = & \delta_{IJ}\delta_{ij} \int_{-\pi /a}^{\pi /a} \frac{\dd^2q}{(2\pi)^2} \, e^{iaq\cdot (x-y)}  
\widehat{\Delta}_{-I}(q), \nn \\
\vev{\Psi_{I\alpha i}(x)\bar{\Psi}_{J\beta j}(y)}_0 & = & \delta_{IJ}\delta_{ij} \int_{-\pi /a}^{\pi /a} \frac{\dd^2q}{(2\pi)^2} \, e^{iaq\cdot (x-y)}  
\left(\widehat{T}_{I}(q)\right)_{\alpha\beta}, 
\quad (\alpha, \beta=1, \cdots, 4) \nn \\
 & & 
\label{prop_lat_pert} 
\eea
with the fermions expressed by 
\beq
\Psi_I(x) = \left(\begin{array}{c}\psi_{+IL}(x) \\ \bar{\psi}_{-IR}(x) \\ \bar{\psi}_{-IL}(x) \\ \psi_{+IR}(x) \end{array}\right), \qquad 
\bar{\Psi}_I(x) =\left( \bar{\psi}_{+IL}(x), \psi_{-IR}(x), \psi_{-IL}(x), \bar{\psi}_{+IR}(x)\right), 
\eeq
and 
\bea
\widehat{\Delta}_{\pm I}(q) & \equiv & \frac{1}{\bar{q}^2 + \left(\frac{ra}{2}\hat{q}^2\right)^2 + \twm_I\twm_{\pm I}^*} \, , \nn \\
\widehat{T}_I(q) & \equiv & \widehat{\Delta}_{+I}(q) \left[ \begin{array}{cccc} 
-i\bar{q}_0-\bar{q}_1 & -\frac{ra}{2}\hat{q}^2 & 0 & -\twm_I \\
 0 & 0 & 0 & 0 \\ 
\frac{1}{\twm_I}\frac{ra}{2}\hat{q}^2 (i\bar{q}_0 +\bar{q}_1) & \frac{1}{\twm_I}\left(\frac{ra}{2}\hat{q}^2\right)^2 & 0 & \frac{ra}{2}\hat{q}^2 \\
-\twm_{+I}^* -\frac{1}{\twm_I}\left(\frac{ra}{2}\hat{q}^2\right)^2 & \frac{1}{\twm_I}\frac{ra}{2}\hat{q}^2 (-i\bar{q}_0 +\bar{q}_1) & 0 & 
-i\bar{q}_0 +\bar{q}_1 
\end{array} \right] \nn \\
 & & \hspace{-0.3cm}+ \widehat{\Delta}_{-I}(q) \left[ \begin{array}{cccc} 
0 & 0 & 0 & 0 \\
-\frac{ra}{2}\hat{q}^2 & -i\bar{q}_0 +\bar{q}_1 &  -\twm_I &  0 \\
-\frac{1}{\twm_I}\frac{ra}{2}\hat{q}^2 (i\bar{q}_0 +\bar{q}_1) & -\twm_{-I}^* -\frac{1}{\twm_I}\left(\frac{ra}{2}\hat{q}^2\right)^2 & 
-i\bar{q}_0 -\bar{q}_1 & 0 \\
\frac{1}{\twm_I}\left(\frac{ra}{2}\hat{q}^2\right)^2 & -\frac{1}{\twm_I}\frac{ra}{2}\hat{q}^2 (-i\bar{q}_0 +\bar{q}_1) & 
\frac{ra}{2}\hat{q}^2 & 0 
\end{array} \right]. \nn \\
 & & 
\eea
Here, we use the notations for the lattice momenta: 
\bea
\bar{q}_\mu \equiv \frac{1}{a}\sin \left(aq_{\mu}\right), & & \bar{q}^2 = \sum_{\mu=0}^1 \bar{q}_{\mu}^2, \nn \\
\hat{q}_\mu \equiv \frac{2}{a} \sin\left(\frac{aq_{\mu}}{2}\right), & & \hat{q}^2 = \sum_{\mu=0}^1 \hat{q}_{\mu}^2. 
\label{lat_mom}
\eea
In the calculation of $-\dvev{\sum_{\mu=0}^1\nabla_{\mu}^* \check{J}_{A\mu}(x)}$ and 
$\dvev{{\cal M}(x)}$, 
where $\dvev{\cdot}$ represents the expectation value with respect to the matter sector under the action 
$S^{\rm LAT}_{{\rm mat}, +\twm} + S^{\rm LAT}_{{\rm mat}, -\twm}$, 
the SYM fields are treated as the external fields. 
To see the anomaly from the lattice theory, we focus on the case that the external momenta of the SYM fields 
are much smaller than the scale $1/a$. The explicit form of the interaction terms 
\beq
  S^{\rm LAT-int}_{{\rm mat}, \twm} = V^{\rm int}_1 + \cdots + V^{\rm int}_7
\eeq  
is given in appendix~\ref{app:lat_pert}, where we keep only the terms with 
\beq
(\text{the power of external momenta}) + (\text{the number of the SYM fields}) \leq 2, 
\eeq
which are relevant for the computation. 
$V^{\rm int}_1$ and $V^{\rm int}_2$ are the three- and four-point gauge-squark couplings\footnote{\label{foot:quark}
Here, we call the charged scalars $\phi_{\pm I}$ and fermions $\psi_{\pm I L}, \psi_{\pm I R}$ as ``squarks'' and ``quarks''.}, and 
$V^{\rm int}_3$ consists of the three-point gauge-quark couplings. 
$V^{\rm int}_4$ contains the three- and four-point interactions of the squarks to the Higgs, $F_{01}$ or 
$D$. 
Also, $V^{\rm int}_5$ consists of the Yukawa coupling of the Higgs to the quarks. 
$V^{\rm int}_6$ and $V^{\rm int}_7$ are the Yukawa couplings containing $\chi, \eta$ and $\psi_{\mu}$, 
respectively. 
 
Next, let us calculate $-\dvev{\sum_{\mu=0}^1\nabla_{\mu}^* \check{J}_{A\mu}(x)}$.  
In the lattice perturbation, it is easy to see that the zeroth order contribution vanishes. 
In the first order, 
\beq
\dvev{\sum_{\mu=0}^1\nabla_{\mu}^* \check{J}_{A\mu}(x) \, V^{\rm int}_s}_{C, 0} \qquad (s=1, \cdots, 5)
\eeq
(with the suffix ``$C, 0$'' meaning to take the connected Feynman diagrams) 
give no contribution within the linear order of the external momenta. 
The contribution from $V^{\rm int}_6$ leads to 
\bea
  & & \dvev{\sum_{\mu=0}^1\nabla_{\mu}^* \check{J}_{A\mu}(x) \, V^{\rm int}_6}_{C, 0} = 
  \sum_{I=1}^n \int \frac{\dd^2k}{(2\pi)^2}\frac{\dd^2k'}{(2\pi)^2} \, e^{ia(k+k')\cdot x} \nn \\
 & & \hspace{6cm} \times \sum_{\mu=0}^1 C_{I \mu}(a) (- k_\mu-k'_\mu)\, \tr \left(\wt{\psi}_\mu(k) \wt{\eta}(k')\right), \\
  & & C_{I\mu}(a)  \equiv \left(\frac{ra}{2}\right)^2 \int^{\pi /a}_{-\pi /a} \frac{\dd^2q}{(2\pi)^2} \, \hq^2 \cos(aq_\mu) 
\left(\widehat{\Delta}_{+I}(q)^2 + \widehat{\Delta}_{-I}(q)^2\right), 
\eea 
which is to be compared with the contributions from the second order perturbation and must be irrelevant to the anomaly. 
The irrelevance can be seen as follows.  
Since $C_{I\mu}(a)$ is evaluated to be 
$\cO(a^2)$ (up to the possible logarithmic factors) 
for fixed $r>0$, the contribution can be neglected in the continuum limit $a\to 0$. 
Similarly, the contribution from $V^{\rm int}_7$ is negligible in the continuum limit. 
Thus, up to the first order perturbation, we conclude that $-\dvev{\sum_{\mu=0}^1\nabla_{\mu}^* \check{J}_{A\mu}(x)}$ 
vanishes in the continuum limit. 
Namely, the contribution to the ${\rm U}(1)_A$ Noether current from the Wilson terms does not lead to the anomaly. 
It is a plausible consequence, since the Wilson terms preserve the ${\rm U}(1)_A$ symmetry\footnote{
This situation is different from that of the chiral anomaly in the Wilson fermions. 
In that case, the Wilson terms in the lattice action break the chiral symmetry, and thus the anomaly 
arises from the $r$-dependent part of the Noether current.}. 

Regarding $\dvev{{\cal M}(x)}$, we easily see that the zeroth order of $\dvev{{\cal M}_F(x)}$ vanishes. 
In the next order, as discussed in the continuum case, we take into account the contributions from  
the zeroth order of $\dvev{{\cal M}_B(x)}$ and from the first order of $\dvev{{\cal M}_F(x)}$. 
The first order contributions of $\dvev{{\cal M}_F(x)}$ with the vertices other than $V^{\rm int}_3$ and $V^{\rm int}_5$ 
trivially vanish, and it turns out that up to the irrelevant pieces
\bea
 -\dvev{{\cal M}_F(x)V^{\rm int}_5}_{C,0} & =  & 2\sum_{I=1}^n\left[-\twm_I \, \tr \,\bar{\phi}(x) 
\left\{L(\twm_I\twm_{+I}^*) + L(\twm_I\twm_{-I}^*)\right\}  \right. \nn \\
 & & \hspace{1cm} \left. + \tr \,\phi(x) \left\{\twm_{+I}^* L(\twm_I\twm_{+I}^*) + \twm_{-I}^* L(\twm_I\twm_{-I}^*)\right\}\right], \\
L(m^2) & \equiv & \int^{\pi /a}_{-\pi /a} \frac{\dd^2q}{(2\pi)^2} \, 
\frac{1}{\bar{q}^2 + \left(\frac{ra}{2}\hat{q}^2\right)^2 + m^2},  
\label{MV5_lat}
\eea
which cancels with the zeroth order of $\dvev{{\cal M}_B(x)}$. 
The remaining $-\dvev{{\cal M}_F(x)V^{\rm int}_3}_{C,0}$ potentially contributes to the anomaly. 
It is computed to obtain 
\bea
 & & -\dvev{{\cal M}_F(x)V^{\rm int}_3}_{C,0} =  \sum_{I=1}^n \widehat{C}_I \, \tr \, F_{01}(x) +\sum_{I=1}^n \check{C}_I \, \tr \, F_{01}(x), 
\label{MV3_lat}\\
 & & \widehat{C}_I \equiv  4\int^{\pi /a}_{-\pi /a} \frac{\dd^2q}{(2\pi)^2} \, \cos (aq_0)\cos(aq_1) 
\left(\twm_I\twm_{+I}^*\widehat{\Delta}_{+I}(q)^2 -\twm_I\twm_{-I}^*\widehat{\Delta}_{-I}(q)^2\right), \nn \\
 & & \check{C}_I = \check{C}_{+I} - \check{C}_{-I}, \\
 & & \check{C}_{\pm I} \equiv  (ra)^2 \int^{\pi /a}_{-\pi /a} \frac{\dd^2q}{(2\pi)^2} \, \hq^2 \left(\hq^2 \cos(aq_0)\cos(aq_1) 
-2\bq_0^2\cos(aq_1)-2\bq_1^2\cos(aq_0)\right) \nn \\ 
& & \hspace{4cm}\times\widehat{\Delta}_{\pm I}(q)^2. 
 \eea
The first term in the r.h.s. of (\ref{MV3_lat}) gives the counterpart of $\dvev{M(x)}$ in the continuum theory. In fact, 
according to the Reisz theorem~\cite{reisz}, since $\widehat{C}_I$ has the lattice degrees of divergence\footnote{
Let us consider some amplitude ${\cal A}$ with $L$ loop momenta $q^{(l)}$ $(l=1, \cdots, L)$. 
Suppose in scaling 
\beq
a \to \frac{1}{\Lambda} a, \qquad q^{(l)} \to \Lambda q^{(l)}, 
\eeq
with $\Lambda$ large, the amplitude ${\cal A}$ behaves as
\beq
{\cal A} = \cO(\Lambda^D). 
\eeq
Then, the lattice degrees of divergence of ${\cal A}$ is $D$.} 
 $-2$, we can naively take the continuum limit to get  
\beq
\widehat{C}_I \to 4\int \frac{\dd^2q}{(2\pi)^2}\, \left[\frac{\twm_I\twm_{+I}^*}{(q^2 +\twm_I\twm_{+I}^*)^2}   
-\frac{\twm_I\twm_{-I}^*}{(q^2 +\twm_I\twm_{-I}^*)^2} \right]. 
\eeq
{}From this, it can be checked that $\sum_{I=1}^n \widehat{C}_I \, \tr \, F_{01}(x)$ coincides to the second term in the r.h.s. 
of (\ref{anomaly_cont_lhs}) up to the linear order of the external momenta for the case $n_+ = n_-$. 
In the second term, the prefactor $(ra)^2$ in $\check{C}_{\pm I}$ indicates that its origin is purely quantum mechanical. 
For the case of all the twisted masses finite, we obtain 
\beq
\check{C}_{+I} = \check{C}_{-I} = -\frac{1}{\pi} 
\eeq
in the continuum limit~\cite{karsten-smit}, and $\sum_{I=1}^n \check{C}_I \, \tr \, F_{01}(x)$ 
coincides to the first term in the r.h.s. of (\ref{anomaly_cont_lhs}) vanishing for $n_+ = n_-$. 

Since the $\cO(a)$ terms in the current $\hat{J}^{\rm mat}_{A\mu}(x)$ turn out to give no 
relevant contribution, the perturbative computation on the lattice leads to 
\beq
\dvev{\der_{\mu} j^{\rm mat}_{A\mu}(x)} = -\frac{1}{\pi}(n-n) \, \tr \, F_{01}(x) 
+ \dvev{M(x)} 
\eeq     
in the continuum limit, which coincides to the continuum result (\ref{anomaly_cont'}) for $n_+ = n_-$.  

\paragraph{Decoupling.} 
So far, we have considered the case $n_+ = n_- (\equiv n)$, 
and the obtained result coincides with the continuum case (\ref{anomaly_cont}).   
If some of the matter multiplets $\Phi_{+I}$'s or $\Phi_{-I}$'s decouple from the theory 
by sending the corresponding anti-holomorphic twisted masses $\twm_{+I}^*$'s or $\twm_{-I}^*$'s to the infinity, 
we can analyze the general situation of $n_+\neq n_-$,   
although the lattice theory is defined only in the case $n_+ = n_-$. 

Let us see whether the decoupling holds in the computation of the anomaly. 
For example, we send $\twm_{-n}^*$ to the infinity before taking the continuum limit $a\to 0$ (equivalently, 
take $\twm_{-n}^*$ much larger than $1/a$ at the lattice level). 
Since the loop momenta run over the finite range $[-\pi /a, \pi /a]$, 
the decoupling can be clearly discussed in contrast to the continuum case. 
In the calculation of this subsection, the decoupling is achieved due to 
\beq
\widehat{\Delta}_{-n}(q) \to 0 \qquad (\twm_{-n}^* \to \infty), 
\eeq
except the term in (\ref{MV5_lat}): $\twm_{-n}^* L(\twm_n\twm_{-n}^*)$. 
But, since (\ref{MV5_lat}) totally cancels with the zeroth order of $\dvev{{\cal M}_B(x)}$, 
it does not appear. It is due to the $Q$-supersymmetry.  
In particular, $\check{C}_I$ in (\ref{MV3_lat}) becomes 
\beq
\check{C}_I = -\frac{1}{\pi} (n-(n-1)), 
\eeq
to produce the correct value of the anomaly. Since the argument is same also for the other twisted masses $\twm_{\pm I}^*$, 
by sending 
\beq
\twm_{+I}^* \to \infty \qquad (I=n_++1, \cdots, n), \qquad \twm_{-I'}^* \to \infty \qquad (I'=n_-+1, \cdots, n),
\eeq
the corresponding fields decouple in the calculation 
and 
we reproduce the anomalous ${\rm U}(1)_A$ WT identity for $n_+$ fundamental and $n_-$ anti-fundamental matters: 
\beq
\vev{\der_\mu j_{A\mu}} = -\frac{1}{\pi}(n_+-n_-) \vev{\tr \, F_{01}(x)} +  \vev{M(x)}
\eeq
in the continuum limit. 
Note that the decoupling is not completely trivial, 
because we send only the anti-holomorphic twisted masses $\twm_{+I}$ or $\twm_{-I}$ infinitely 
massive, while the holomorphic twisted masses $\twm_{+I}=\twm_{-I}(=\twm_I)$ are kept finite. 
In this case, the $Q$-supersymmetry plays an important role for the decoupling.

\setcounter{equation}{0}
\section{Summary and Discussion}
\label{sec:summary}
In this paper, we have discussed the lattice formulation of two-dimensional $\cN=(2,2)$ SQCD with 
$n_+$ fundamental and $n_-$ anti-fundamental matters, preserving the supercharge $Q$ exactly. 

We introduced the Wilson terms to suppress the species doublers of the matter fields, but then it was necessary to 
take $n_+=n_-(\equiv n)$, as long as respecting the gauge symmetry and the $Q$-supersymmetry. 
When introducing the twisted mass terms into the theory, the $Q$-supersymmetry transformation is deformed so that 
its nilpotency holds up to the combination of an infinitesimal gauge transformation and infinitesimal flavor rotations. 
The transformation parameters are the Higgs scalar $\phi(x)$ and the holomorphic twisted masses 
$\twm_{\pm I}$ ($I=1, \cdots, n$), respectively. 
Differently from the continuum case, we focused on the case $\twm_{+I} =\twm_{-I}(\equiv \twm_I)$ so that 
the $Q$-nilpotency entirely holds on the lattice and the $Q$-exact lattice action is guaranteed to be $Q$-invariant. 
It is due to the Wilson terms that reduce the flavor symmetry ${\rm U}(1)^n \times {\rm U}(1)^n$ to its diagonal subgroup. 

Although the $Q$-invariant lattice action is applicable to the case $n_+=n_-$, 
if some of the fundamental or anti-fundamental multiplets decouple from the theory 
by sending the corresponding anti-holomorphic twisted masses ($\twm_{+I}^*$'s or $\twm_{-I}^*$'s) to the infinity, 
we can analyze the general case $n_+\neq n_-$ starting from the lattice action. 
In fact, we have shown it possible in computing the ${\rm U}(1)_A$ anomaly by the lattice perturbation. 
The decoupling is not a trivial consequence, because the holomorphic twisted masses $\twm_I$ are kept finite. 
It should be noted that the $Q$-supersymmetry plays an important role to achieve the decoupling. 
When considering other observables, even in the situation that the decoupling does not hold completely, 
we could analyze the case $n_+\neq n_-$ by adding appropriate 
counter terms to the lattice action. 

It is certainly desirable to construct the $Q$-invariant lattice action for the general $n_{\pm}$. 
Through our construction of the action, the problem is seen closely related to the realization of the chiral (flavor) symmetry 
of the lattice action, which is explicitly broken by the Wilson terms in our construction. 
It would be a crucial step to improve our action to use the Ginsparg-Wilson fermions~\cite{ginsparg-wilson} for the matter sector 
with maintaining the exact $Q$-supersymmetry. 
In the two-dimensional Wess-Zumino model, the Ginsparg-Wilson fermions are introduced to the lattice formulation 
with the exact supersymmetry by using the Nicolai mapping~\cite{kikukawa-nakayama}. 
It would give a hint to construct our desirable lattice action.  
As discussed in \cite{kikukawa-nakayama}, such construction leads to the exactly holomorphic 
or anti-holomorphic superpotential terms, which further help decreasing the number of the relevant operators to be tuned.   

The two-dimensional $\cN =(2,2)$ SQCD models with various superpotentials have been analytically investigated based on 
the effective twisted superpotentials~\cite{witten2,hanany-hori,hori-tong}. 
The number of the vacua or the Witten index of the models has been computed for various $N, n_{\pm}$, 
and the analog of the Seiberg duality in four dimensions has been discussed. 
Some insights have been obtained with respect to the property of the sigma models on Calabi-Yau manifolds via the correspondence 
between the gauged linear sigma models and the nonlinear sigma models, where the $D$-term condition in the former determines 
the target space of the latter in the infra-red limit. 
It will be worth confirming those properties and exploring new aspects, which are not yet investigated there,  
from the first principle computation using the lattice formulation.

\vspace{1cm}
{\bf Acknowledgments}\\
The author would like to thank Poul~H.~Damgaard, 
Issaku~Kanamori, Yoshio~Kikukawa, So~Matsuura, Hiroshi~Suzuki, Tomohisa~Takimi and Piljin~Yi  
for valuable discussions. 
Also, he would like to express his gratitude to the Niels Bohr Institute for hospitality during his 
visit when a part of this work was done. 

\vspace{1cm}

\appendix
\section{Continuum $(1+1)$-Dimensional ${\cal N}= (2,2)$ SQCD}
\label{app:cont(1+1)DSQCD}
\setcounter{equation}{0}
First, we start from $\cN =1$ SQCD with $n_+$ fundamental and 
$n_-$ anti-fundamental matters in $(3+1)$-dimensional Minkowski space $x'^m$ $(m=0, \cdots, 3)$, 
to give
the corresponding $(1+1)$-dimensional $\cN = (2,2)$ SQCD, via the dimensional reduction.
The action of the $(3+1)$-dimensional $\cN =1$ SQCD is expressed in terms of the $\cN =1$ superfields as\footnote{We use 
the notation of Wess-Bagger's book~\cite{WB}.}  
\bea
S_{\rm 4DSQCD} & = & \int \dd^4x' \left[ \frac{1}{8g^2} \tr \left( \thth{W^\alpha W_\alpha}
+ \tbtb{\bar{W}_{\dot{\alpha}} \bar{W}^{\dot{\alpha}}}\right)\right. \nn \\ 
 & &  \hspace{1cm}\left. +\sum_{I=1}^{n_+}\tttt{\Phi_{+I}^\dagger \, e^V \, \Phi_{+I}} 
+ \sum_{I'=1}^{n_-}\tttt{\Phi_{-I'} \, e^{-V} \, \Phi_{-I'}^\dagger} \right], 
\eea
where $V$ is a vector superfield, and $\Phi_{+I}$ ($\Phi_{-I'}$) are chiral superfields 
belonging to the fundamental (anti-fundamental) representation of the gauge group $G$, i.e. 
column (row) vectors. 
Under the gauge transformation with the parameter $\Lambda$ being a chiral superfield, 
they transform as 
\bea
 e^V \to e^{-i\Lambda^\dagger} \, e^V \, e^{i\Lambda}, &  & e^{-V} \to e^{-i\Lambda} \, e^{-V} \, e^{i\Lambda^\dagger}, \nn \\
 \Phi_{+I} \to e^{-i\Lambda} \, \Phi_{+I}, &  & \Phi_{-I'} \to \Phi_{-I'} \, e^{i\Lambda}, \nn \\
 \Phi_{+I}^\dagger \to \Phi_{+I}^\dagger \, e^{i\Lambda^\dagger}, & & 
\Phi_{-I'}^\dagger \to e^{-i\Lambda^\dagger} \, \Phi_{-I'}^\dagger. 
\eea
  
After taking the Wess-Zumino gauge, the action is written in terms of the component fields as
\bea
S_{\rm 4DSQCD} & = & \int \dd^4x' \left[ \frac{2}{g^2} \, \tr \left(-\frac14F^{mn}F_{mn} -i\bar{\lambda}\bar{\sigma}^m\cD_m\lambda 
+\frac12 D^2\right) \right. \nn \\
 & & +\sum_{I=1}^{n_+}\left( -\cD^m\phi_{+I}^\dagger \cD_m\phi_{+I}
 +\phi_{+I}^\dagger D\phi_{+I} +F_+^\dagger F_+ \right. \nn \\
 & & \hspace{1.5cm} \left. -i\bar{\psi}_{+I}\bar{\sigma}^m\cD_m\psi_{+I} +i\sqrt{2}\left(\phi_{+I}^\dagger\lambda\psi_{+I} 
- \bar{\psi}_{+I}\bar{\lambda}\phi_{+I}\right)\right)  \nn \\
 & & +\sum_{I'=1}^{n_-}\left(-\cD^m\phi_{-I'}\cD_m\phi_{-I'}^\dagger  -\phi_{-I'}D\phi_{-I'}^\dagger + F_-F_-^\dagger \right.\nn \\
  & & \left.\left.\hspace{1.5cm} -i\psi_{-I'}\sigma^m\cD_m\bar{\psi}_{-I'} +i\sqrt{2}\left(-\psi_{-I'}\lambda\phi_{-I'}^\dagger 
+\phi_{-I'}\bar{\lambda}\bar{\psi}_{-I'}\right)\right)  \right]. 
\eea
Here, $v_m, \lambda, D$ are components of $V$, and $\phi_\pm, \psi_\pm, F_\pm$ are of $\Phi_\pm$. 
We rescaled as $v_m \to 2 v_m$, $\lambda \to 2\lambda$, $D\to 2D$ in Wess-Bagger's notation, i.e. 
\beq
V = -2\theta\sigma^m\bar{\theta} \, v_m(x') +2i\theta\theta\bar{\theta}\bar{\lambda}(x')-2i\bar{\theta}\bar{\theta}\theta\lambda(x') 
+\theta\theta\bar{\theta}\bar{\theta}D(x'), 
\eeq
then the field strength and the covariant derivatives are expressed by ($\der_m \equiv \der/\der x'^m$) 
\bea
 F_{mn} = \der_mv_n-\der_nv_m+i[v_m, v_n], & &  \cD_m\lambda = \der_m\lambda +i[v_m, \lambda], \nn \\
 \cD_m\phi_{+I} = \der_m\phi_{+I} +iv_m \, \phi_{+I}, & &  \cD_m\psi_{+I} = \der_m\psi_{+I} + iv_m \, \psi_{+I}, \nn \\
 \cD_m\phi_{+I}^\dagger = \der_m\phi_{+I}^\dagger -i\phi_{+I}^\dagger \, v_m, & & 
\cD_m \bar{\psi}_{+I} = \der_m\bar{\psi}_{+I} -i\bar{\psi}_{+I} \, v_m \nn \\
\cD_m\phi_{-I'} = \der_m\phi_{-I'} -i\phi_{-I'} \, v_m, & &  \cD_m\psi_{-I'}= \der_m\psi_{-I'} -i\psi_{-I'} \, v_m, \nn \\
 \cD_m\phi_{-I'}^\dagger = \der_m\phi_{-I'}^\dagger +iv_m \, \phi_{-I'}^\dagger, & & 
\cD_m\bar{\psi}_{-I'} = \der_m \bar{\psi}_{-I'} + iv_m \, \bar{\psi}_{-I'}.
\eea
The supersymmetry transformation, which keeps the Wess-Zumino gauge, is given by 
\bea
\delta_\xi V & = & (\xi Q + \bar{\xi}\bar{Q})V + \delta_\Lambda V, \nn \\
\delta_\xi \Phi_{+I} & = & (\xi Q + \bar{\xi}\bar{Q}) \Phi_{+I} +\delta_\Lambda \Phi_{+I}, \nn \\
\delta_\xi \Phi_{-I'} & = & (\xi Q + \bar{\xi}\bar{Q}) \Phi_{-I'} + \delta_\Lambda \Phi_{-I'}, 
\label{SUSY_WZ}
\eea
where 
\beq
Q_\alpha = \frac{\der}{\der \theta^\alpha}-i\sigma^m_{\alpha\dot{\alpha}}\bar{\theta}^{\dot{\alpha}}\der_m, \qquad 
\bar{Q}_{\dot{\alpha}} = -\frac{\der}{\der \bar{\theta}^{\dot{\alpha}}} +i\theta^\alpha\sigma^m_{\alpha\dot{\alpha}}\der_m,
\eeq
and $\xi, \bar{\xi}$ are spinor parameters of the transformation. 
Since the tranformation $(\xi Q + \bar{\xi}\bar{Q})$ alone does not preserve the Wess-Zumino gauge, 
the infinitesimal super gauge transformations of the last terms in the r.h.s. of (\ref{SUSY_WZ}) are necessary 
to recover the Wess-Zumino gauge: 
\bea
 & & \delta_\Lambda V   =  2i(\Lambda -\Lambda^\dagger) +i [V, \Lambda + \Lambda^\dagger], \qquad 
\delta_\Lambda \Phi_{+I} = -2i\Lambda\Phi_{+I}, \qquad \delta_\Lambda\Phi_{-I'} = 2i\Phi_{-I'}\Lambda, \nn \\
 & & \Lambda = -i\theta\sigma^m\bar{\xi} \, v_m(x') -\theta\theta\bar{\xi}\bar{\lambda}(x') 
+\frac12\theta\theta\bar{\theta}\bar{\sigma}^n\sigma^m\bar{\xi} \, \der_nv_m(x'), \nn \\
 & & \Lambda^\dagger = -i\bar{\theta}\bar{\sigma}^m\xi \, v_m(x') -\bar{\theta}\bar{\theta}\xi\lambda(x') 
+\frac12\bar{\theta}\bar{\theta}\theta\sigma^n\bar{\sigma}^m\bar{\xi} \, \der_nv_m(x'). 
\eea 
 
Next, we collapse the directions $x'^1, x'^2$ to points, and denote 
\beq
x^0 \equiv x'^0, \quad x^1 \equiv x'^3, \qquad A_0 \equiv v_0, \quad A_1 \equiv v_3, \qquad X_1 \equiv v_1, \quad X_2\equiv v_2
\eeq
to obtain the $(1+1)$-dimensional $\cN =(2,2)$ SQCD: 
\bea
S_{\rm 2DSQCD} & = & S_{\rm SYM} + S_{{\rm mat}, +} + S_{{\rm mat}, -} \, , \\
S_{\rm SYM} & = & \frac{2}{g^2} \int \dd^2x \,\tr \left(-\frac14F^{\mu\nu}F_{\mu\nu} -\frac12\cD^\mu\phi\cD_\mu\bar{\phi} 
-\frac18[\phi, \bar{\phi}]^2+\frac12 D^2  \right. \nn \\
 & & \hspace{2.5cm} \left. -i\bar{\lambda}\bar{\sigma}^\mu\cD_\mu\lambda 
-\bar{\lambda}_{\dot{1}}[\bar{\phi}, \lambda_2]-\bar{\lambda}_{\dot{2}}[\phi,\lambda_1]\frac{}{}\right),  \nn\\
S_{{\rm mat}, +}  &= &  \int \dd^2x\sum_{I=1}^{n_+} \left[
 -\cD^\mu\phi_{+I}^\dagger\cD_\mu\phi_{+I} -\frac12\phi_{+I}^\dagger\{\phi, \bar{\phi}\}\phi_{+I} + F_{+I}^\dagger F_{+I} 
+ \phi_{+I}^\dagger D\phi_{+I} \right.\nn \\
 & & \hspace{2cm} -i\bar{\psi}_{+I}\bar{\sigma}^\mu\cD_\mu\psi_{+I} 
-\bar{\psi}_{+I\dot{1}}\bar{\phi} \, \psi_{+I2}
-\bar{\psi}_{+I\dot{2}}\phi \, \psi_{+I1} \nn \\
 & & \hspace{2cm} \left. +i\sqrt{2}\left(\phi_{+I}^\dagger\lambda\psi_{+I} -\bar{\psi}_{+I}\bar{\lambda}\phi_{+I}\right)\right],  \nn\\
S_{{\rm mat}, -} & =&   \int \dd^2x \sum_{I'=1}^{n_-}\left[
-\cD^\mu\phi_{-I'}\cD_\mu\phi_{-I'}^\dagger -\frac12\phi_{-I'}\{\phi, \bar{\phi}\}\phi_{-I'}^\dagger + F_{-I'}F_{-I'}^\dagger  
- \phi_{-I'} D\phi_{-I'}^\dagger \right. \nn \\
 & & \hspace{2cm} -i\psi_{-I'}\sigma^\mu\cD_\mu\bar{\psi}_{-I'} 
-\psi_{-I'2}\bar{\phi} \, \bar{\psi}_{-I'\dot{1}}
-\psi_{-I'1}\phi \, \bar{\psi}_{-I'\dot{2}} \nn \\
 & & \hspace{2cm} \left. +i\sqrt{2}\left(-\psi_{-I'}\lambda\phi_{-I'}^\dagger +\phi_{-I'}\bar{\lambda}\bar{\psi}_{-I'}\right) \right],  \nn  
\eea
where $\der_\mu\equiv \der/\der x^\mu$ ($\mu =0, 1$), 
$\cD_\mu$ are the covariant derivatives with the gauge fields $A_\mu$ used, 
and $\phi=X_1+iX_2$, $\bar{\phi}=X_1-iX_2$ are complex Higgs scalars.  

Using the index $R$ ($L$) instead of the spinor index $1$ ($2$) to show the spinor structure explicitly, we have 
\bea
S_{\rm SYM} & = &  \frac{1}{g^2} \int \dd^2x \,\tr \left(-\frac12F^{\mu\nu}F_{\mu\nu} -\cD^\mu\phi\cD_\mu\bar{\phi} 
-\frac14[\phi, \bar{\phi}]^2+D^2  \right. \nn \\
 & & \hspace{2.5cm} \left. +2i\bar{\lambda}_R\cD_L\lambda_R +2i\bar{\lambda}_L\cD_R\lambda_L 
-\bar{\lambda}_R[\bar{\phi}, \lambda_L]-\bar{\lambda}_L[\phi,\lambda_R]\frac{}{}\right),  \nn\\
S_{{\rm mat}, +}  &= &  \int \dd^2x \sum_{I=1}^{n_+}\left[
 -\cD^\mu\phi_{+I}^\dagger\cD_\mu\phi_{+I} -\frac12\phi_{+I}^\dagger\{\phi, \bar{\phi}\}\phi_{+I} + F_{+I}^\dagger F_{+I} 
+ \phi_{+I}^\dagger D\phi_{+I} \right.\nn \\
 & & \hspace{1cm} +2i\bar{\psi}_{+IR}\cD_L\psi_{+IR} +2i\bar{\psi}_{+IL}\cD_R\psi_{+IL} 
-\bar{\psi}_{+IR} \, \bar{\phi} \, \psi_{+IL} -\bar{\psi}_{+IL} \, \phi \, \psi_{+IR} \nn \\
 & & \hspace{1cm} \left. +i\sqrt{2}\left(\phi_{+I}^\dagger(\lambda_L\psi_{+IR}-\lambda_R\psi_{+IL}) 
+(-\bar{\psi}_{+IR}\bar{\lambda}_L + \bar{\psi}_{+IL}\bar{\lambda}_R)\phi_{+I}\right)\right] , \nn\\
S_{{\rm mat}, -} & =&   \int \dd^2x \sum_{I'=1}^{n_-}\left[
-\cD^\mu\phi_{-I'}\cD_\mu\phi_{-I'}^\dagger -\frac12\phi_{-I'}\{\phi, \bar{\phi}\}\phi_{-I'}^\dagger + F_{-I'}F_{-I'}^\dagger  
- \phi_{-I'} D\phi_{-I'}^\dagger \right. \nn \\
 & & \hspace{1cm} +2i\psi_{-I'R}\cD_L\bar{\psi}_{-I'R} +2i\psi_{-I'L}\cD_R\bar{\psi}_{-I'L} 
-\psi_{-I'L} \, \bar{\phi} \, \bar{\psi}_{-I'R} -\psi_{-I'R} \, \phi \, \bar{\psi}_{-I'L}  \nn \\
  & & \hspace{1cm} \left. +i\sqrt{2}\left((-\psi_{-I'L}\lambda_R + \psi_{-I'R}\lambda_L)\phi_{-I'}^\dagger 
+\phi_{-I'}(\bar{\lambda}_R\bar{\psi}_{-I'L} -
\bar{\lambda}_L\bar{\psi}_{-I'R})\right) \right]  \nn \\
 & &
\label{2dSQCD_Min}
\eea
with $\cD_R \equiv \frac12 (\cD_0-\cD_1)$, $\cD_L \equiv \frac12 (\cD_0 + \cD_1)$. 
Correspondingly, the supersymmetry transformations for the component fields are written as 
\bea
\delta_\xi A_0 & = & i\xi_L\bar{\lambda}_L + i\xi_R\bar{\lambda}_R + i\bar{\xi}_R\lambda_R +i\bar{\xi}_L\lambda_L, \nn \\
\delta_\xi A_1 & = & i\xi_L\bar{\lambda}_L - i\xi_R\bar{\lambda}_R - i\bar{\xi}_R\lambda_R +i\bar{\xi}_L\lambda_L, \nn \\
\delta_\xi \phi & = & -2i\xi_L\bar{\lambda}_R -2i\bar{\xi}_R\lambda_L, \nn \\
\delta_\xi \bar{\phi} & = & -2i\xi_R\bar{\lambda}_L -2i\bar{\xi}_L\lambda_R, \nn \\
\delta_\xi \lambda_R & = & \xi_R\left(iD + F_{01} +\frac{i}{2}[\phi, \bar{\phi}]\right) +2\xi_L\cD_R\bar{\phi}, \nn \\
\delta_\xi \lambda_L & = & \xi_L\left(iD - F_{01} -\frac{i}{2}[\phi, \bar{\phi}]\right) +2\xi_R\cD_L\phi, \nn \\
\delta_\xi \bar{\lambda}_R & = & \bar{\xi}_R\left(-iD+F_{01}-\frac{i}{2}[\phi, \bar{\phi}]\right) +2\bar{\xi}_L\cD_R\phi, \nn \\
\delta_\xi \bar{\lambda}_L & = & \bar{\xi}_L\left(-iD-F_{01}+\frac{i}{2}[\phi, \bar{\phi}]\right) +2\bar{\xi}_R\cD_L\bar{\phi}, \nn \\
\delta_\xi D & = & 
-2\bar{\xi}_R\cD_L\lambda_R -2\bar{\xi}_L\cD_R\lambda_L +2\xi_L\cD_R\bar{\lambda}_L +2\xi_R\cD_L\bar{\lambda}_R \nn \\
 & & -i\bar{\xi}_R[\bar{\phi}, \lambda_L] -i\bar{\xi}_L[\phi, \lambda_R] +i\xi_L[\bar{\phi}, \bar{\lambda}_R] +i\xi_R[\phi, \bar{\lambda}_L], 
\label{2dsusy_SYM}
\eea
\bea
 \delta_\xi \phi_{+I} & = & \sqrt{2} \, \xi_L\psi_{+I R} -\sqrt{2} \, \xi_R\psi_{+I L}, \nn \\
 \delta_\xi \psi_{+I R} & = & -i2\sqrt{2} \, \bar{\xi}_L\cD_R\phi_{+I} +\sqrt{2} \, \bar{\xi}_R\bar{\phi} \, \phi_{+I} 
+\sqrt{2} \, \xi_R F_{+I}, \nn \\
 \delta_\xi \psi_{+IL} & = & i2\sqrt{2} \, \bar{\xi}_R\cD_L\phi_{+I} -\sqrt{2} \, \bar{\xi}_L\phi \, \phi_{+I} +\sqrt{2} \, \xi_L F_{+I}, \nn \\
 \delta_\xi F_{+I} & = & -i2\sqrt{2} \, \bar{\xi}_R\cD_L\psi_{+IR} -i2\sqrt{2} \, \bar{\xi}_L\cD_R\psi_{+IL} 
 +\sqrt{2} \, \bar{\xi}_R\bar{\phi} \, \psi_{+IL} +\sqrt{2} \, \bar{\xi}_L\phi \, \psi_{+IR} \nn \\
 & & +2i\bar{\xi}_R\bar{\lambda}_L\phi_{+I} -2i\bar{\xi}_L\bar{\lambda}_R\phi_{+I}, \nn \\
\delta_\xi \phi_{+I}^\dagger & = & -\sqrt{2} \, \bar{\xi}_L\bar{\psi}_{+IR} + \sqrt{2} \, \bar{\xi}_R\bar{\psi}_{+IL}, \nn \\
\delta_\xi \bar{\psi}_{+IR} & = & i2\sqrt{2} \, \xi_L\cD_R\phi_{+I}^\dagger +\sqrt{2} \, \xi_R\phi_{+I}^\dagger\phi 
+\sqrt{2} \, \bar{\xi}_R F_{+I}^\dagger, \nn \\
\delta_\xi \bar{\psi}_{+IL} & = & -i2\sqrt{2} \, \xi_R\cD_L\phi_{+I}^\dagger -\sqrt{2} \, \xi_L\phi_{+I}^\dagger\bar{\phi} 
+\sqrt{2} \, \bar{\xi}_L F_{+I}^\dagger, \nn \\
\delta_\xi F_{+I}^\dagger & = & -i2\sqrt{2} \, \xi_L\cD_R\bar{\psi}_{+IL} -i2\sqrt{2} \, \xi_R\cD_L\bar{\psi}_{+IR} 
-\sqrt{2} \, \xi_L\bar{\psi}_{+IR}\bar{\phi} -\sqrt{2} \, \xi_R\bar{\psi}_{+IL}\phi \nn \\
 & & -2i\xi_L\phi_{+I}^\dagger\lambda_R +2i\xi_R\phi_{+I}^\dagger\lambda_L, 
\label{2dsusy_mat+}
\eea
\bea 
 \delta_\xi \phi_{-I'} & = & \sqrt{2} \, \xi_L\psi_{-I' R} -\sqrt{2} \, \xi_R\psi_{-I'L}, \nn \\
\delta_\xi \psi_{-I'R} & = & -i2\sqrt{2} \, \bar{\xi}_L\cD_R\phi_{-I'} -\sqrt{2} \, \bar{\xi}_R\phi_{-I'}\bar{\phi} 
+\sqrt{2} \, \xi_R F_{-I'}, \nn \\
\delta_\xi \psi_{-I'L} & = & i2\sqrt{2} \, \bar{\xi}_R\cD_L\phi_{-I'} +\sqrt{2} \, \bar{\xi}_L\phi_{-I'}\phi +\sqrt{2} \, \xi_L F_{-I'}, \nn \\
\delta_\xi F_{-I'} & = & -i2\sqrt{2} \, \bar{\xi}_R\cD_L\psi_{-I'R} -i2\sqrt{2} \, \bar{\xi}_L\cD_R\psi_{-I'L} 
-\sqrt{2} \, \bar{\xi}_R\psi_{-I'L}\bar{\phi} -\sqrt{2} \, \bar{\xi}_L\psi_{-I'R}\phi \nn \\
 & & -2i\bar{\xi}_R\phi_{-I'}\bar{\lambda}_L +2i\bar{\xi}_L\phi_{-I'}\bar{\lambda}_R, \nn \\
\delta_\xi \phi_{-I'}^\dagger & = & -\sqrt{2} \, \bar{\xi}_L\bar{\psi}_{-I'R} + \sqrt{2} \, \bar{\xi}_R\bar{\psi}_{-I'L}, \nn \\
\delta_\xi \bar{\psi}_{-I'R} & = & i2\sqrt{2} \, \xi_L\cD_R\phi_{-I'}^\dagger -\sqrt{2} \, \xi_R\phi \, \phi_{-I'}^\dagger 
+\sqrt{2} \, \bar{\xi}_R F_{-I'}^\dagger, \nn \\ 
\delta_\xi \bar{\psi}_{-I'L} & = & -i2\sqrt{2} \, \xi_R\cD_L\phi_{-I'}^\dagger +\sqrt{2} \, \xi_L\bar{\phi} \, \phi_{-I'}^\dagger 
+\sqrt{2} \, \bar{\xi}_L F_{-I'}^\dagger, \nn \\ 
\delta_\xi F_{-I'}^\dagger & = & -i2\sqrt{2} \, \xi_L\cD_R\bar{\psi}_{-I'L} -i2\sqrt{2} \, \xi_R\cD_L\bar{\psi}_{-I'R} 
+\sqrt{2} \, \xi_R\phi \, \bar{\psi}_{-I'L} +\sqrt{2} \, \xi_L\bar{\phi} \, \bar{\psi}_{-I'R} \nn \\
 & & +2i\xi_L\lambda_R\phi_{-I'}^\dagger -2i\xi_R\lambda_L\phi_{-I'}^\dagger. 
\label{2dsusy_mat-} 
\eea
 
\subsection{Twisted Masses}
We can introduce the twisted mass terms to the matter multiplets 
by gauging the ${\rm U}(1)^{n_+}\times {\rm U}(1)^{n_-}$ of the flavor symmetry 
and fixing the corresponding vector superfields to the background values as 
\bea
\sum_{I=1}^{n_+}\Phi_{+I}^\dagger \, e^{V} \, \Phi_{+I} & \to & 
\sum_{I=1}^{n_+}\Phi_{+I}^\dagger \, e^{V -\widetilde{V}_{+I}} \, \Phi_{+I}, \nn \\
\sum_{I'=1}^{n_-} \Phi_{-I'} \, e^{-V} \, \Phi_{-I'}^\dagger & \to & 
\sum_{I'=1}^{n_-} \Phi_{-I'} \, e^{-V + \widetilde{V}_{-I'}} \, \Phi_{-I'}^\dagger
\eea
with 
\bea
\widetilde{V}_{+I} & \equiv & 2\theta_R\bar{\theta}_L \, \twm_{+I} +2\theta_L\bar{\theta}_R \, \twm_{+I}^* \, , \nn \\
\widetilde{V}_{-I'} & \equiv & 2\theta_R\bar{\theta}_L \, \twm_{-I'} +2\theta_L\bar{\theta}_R \, \twm_{-I'}^* \, . 
\eea
In the presence of the twisted masses, the matter-part actions $S_{{\rm mat}, +}, S_{{\rm mat}, -}$ are deformed as 
\bea
S_{{\rm mat}, +\twm} & = & \int \dd^2x \sum_{I=1}^{n_+}\left[
 -\cD^\mu\phi_{+I}^\dagger\cD_\mu\phi_{+I} -\frac12\phi_{+I}^\dagger\{\phi -\twm_{+I}, \bar{\phi} -\twm_{+I}^*\}\phi_{+I} \right.\nn \\
 & & \hspace{1cm} + F_{+I}^\dagger F_{+I} + \phi_{+I}^\dagger D\phi_{+I} 
+2i\bar{\psi}_{+IR}\cD_L\psi_{+IR} +2i\bar{\psi}_{+IL}\cD_R\psi_{+IL} \nn \\
 & & \hspace{1cm} -\bar{\psi}_{+IR}(\bar{\phi}-\twm_{+I}^*)\psi_{+IL} -\bar{\psi}_{+IL}(\phi-\twm_{+I})\psi_{+IR} \nn \\
 & & \hspace{1cm} \left. +i\sqrt{2}\left(\phi_{+I}^\dagger(\lambda_L\psi_{+IR}-\lambda_R\psi_{+IL}) 
+(-\bar{\psi}_{+IR}\bar{\lambda}_L + \bar{\psi}_{+IL}\bar{\lambda}_R)\phi_{+I}\right)\right] , \nn \\
\label{2d_S_mat_twm+}\\
S_{{\rm mat}, -\twm} & =&   \int \dd^2x \sum_{I'=1}^{n_-}\left[
-\cD^\mu\phi_{-I'}\cD_\mu\phi_{-I'}^\dagger -\frac12\phi_{-I'}\{\phi-\twm_{-I'}, \bar{\phi}-\twm_{-I'}^*\}\phi_{-I'}^\dagger \right. \nn \\
 & & \hspace{1cm} + F_{-I'}F_{-I'}^\dagger  - \phi_{-I'} D\phi_{-I'}^\dagger 
+2i\psi_{-I'R}\cD_L\bar{\psi}_{-I'R} +2i\psi_{-I'L}\cD_R\bar{\psi}_{-I'L} \nn \\
 & & \hspace{1cm} -\psi_{-I'L}(\bar{\phi}-\twm_{-I'}^*)\bar{\psi}_{-I'R} -\psi_{-I'R}(\phi-\twm_{-I'})\bar{\psi}_{-I'L}  \nn \\
  & & \hspace{1cm} \left. +i\sqrt{2}\left((-\psi_{-I'L}\lambda_R + \psi_{-I'R}\lambda_L)\phi_{-I'}^\dagger 
+\phi_{-I'}(\bar{\lambda}_R\bar{\psi}_{-I'L} -
\bar{\lambda}_L\bar{\psi}_{-I'R})\right) \right] \, .   \nn \\
 & &
\label{2d_S_mat_twm-} 
\eea
Also, the supersymmetry transformations (\ref{2dsusy_mat+}), (\ref{2dsusy_mat-}) become
\bea
 \delta_\xi \phi_{+I} & = & \sqrt{2} \, \xi_L\psi_{+I R} -\sqrt{2} \, \xi_R\psi_{+I L}, \nn \\
 \delta_\xi \psi_{+I R} & = & -i2\sqrt{2} \, \bar{\xi}_L\cD_R\phi_{+I} +\sqrt{2} \, \bar{\xi}_R(\bar{\phi}-\twm_{+I}^*)\phi_{+I} 
+\sqrt{2} \, \xi_R F_{+I}, \nn \\
 \delta_\xi \psi_{+IL} & = & i2\sqrt{2} \, \bar{\xi}_R\cD_L\phi_{+I} -\sqrt{2} \, \bar{\xi}_L(\phi-\twm_{+I})\phi_{+I} 
+\sqrt{2} \, \xi_L F_{+I}, \nn \\
 \delta_\xi F_{+I} & = & -i2\sqrt{2} \, \bar{\xi}_R\cD_L\psi_{+IR} -i2\sqrt{2} \, \bar{\xi}_L\cD_R\psi_{+IL}  \nn \\
 & & +\sqrt{2} \, \bar{\xi}_R(\bar{\phi}-\twm_{+I}^*)\psi_{+IL} +\sqrt{2} \, \bar{\xi}_L(\phi-\twm_{+I})\psi_{+IR} \nn \\
 & & +2i\bar{\xi}_R\bar{\lambda}_L\phi_{+I} -2i\bar{\xi}_L\bar{\lambda}_R\phi_{+I}, \nn \\
\delta_\xi \phi_{+I}^\dagger & = & -\sqrt{2} \, \bar{\xi}_L\bar{\psi}_{+IR} + \sqrt{2} \, \bar{\xi}_R\bar{\psi}_{+IL}, \nn \\
\delta_\xi \bar{\psi}_{+IR} & = & i2\sqrt{2} \, \xi_L\cD_R\phi_{+I}^\dagger +\sqrt{2} \, \xi_R\phi_{+I}^\dagger(\phi-\twm_{+I}) 
+\sqrt{2} \, \bar{\xi}_R F_{+I}^\dagger, \nn \\
\delta_\xi \bar{\psi}_{+IL} & = & -i2\sqrt{2} \, \xi_R\cD_L\phi_{+I}^\dagger -\sqrt{2} \, \xi_L\phi_{+I}^\dagger(\bar{\phi}-\twm_{+I}^*) 
+\sqrt{2} \, \bar{\xi}_L F_{+I}^\dagger, \nn \\
\delta_\xi F_{+I}^\dagger & = & -i2\sqrt{2} \, \xi_L\cD_R\bar{\psi}_{+IL} -i2\sqrt{2} \, \xi_R\cD_L\bar{\psi}_{+IR}  \nn \\
 & & -\sqrt{2} \, \xi_L\bar{\psi}_{+IR}(\bar{\phi}-\twm_{+I}^*) -\sqrt{2} \, \xi_R\bar{\psi}_{+IL}(\phi-\twm_{+I}) \nn \\
 & & -2i\xi_L\phi_{+I}^\dagger\lambda_R +2i\xi_R\phi_{+I}^\dagger\lambda_L, 
\label{2dsusy_mat+_twm}
\eea
\bea 
 \delta_\xi \phi_{-I'} & = & \sqrt{2} \, \xi_L\psi_{-I' R} -\sqrt{2} \, \xi_R\psi_{-I'L}, \nn \\
\delta_\xi \psi_{-I'R} & = & -i2\sqrt{2} \, \bar{\xi}_L\cD_R\phi_{-I'} -\sqrt{2} \, \bar{\xi}_R\phi_{-I'}(\bar{\phi}-\twm_{-I'}^*) 
+\sqrt{2} \, \xi_R F_{-I'}, \nn \\
\delta_\xi \psi_{-I'L} & = & i2\sqrt{2} \, \bar{\xi}_R\cD_L\phi_{-I'} +\sqrt{2} \, \bar{\xi}_L\phi_{-I'}(\phi-\twm_{-I'}) 
+\sqrt{2} \, \xi_L F_{-I'}, \nn \\
\delta_\xi F_{-I'} & = & -i2\sqrt{2} \, \bar{\xi}_R\cD_L\psi_{-I'R} -i2\sqrt{2} \, \bar{\xi}_L\cD_R\psi_{-I'L}  \nn \\
 & & -\sqrt{2} \, \bar{\xi}_R\psi_{-I'L}(\bar{\phi}-\twm_{-I'}^*) -\sqrt{2} \, \bar{\xi}_L\psi_{-I'R}(\phi-\twm_{-I'}) \nn \\
 & & -2i\bar{\xi}_R\phi_{-I'}\bar{\lambda}_L +2i\bar{\xi}_L\phi_{-I'}\bar{\lambda}_R, \nn \\
\delta_\xi \phi_{-I'}^\dagger & = & -\sqrt{2} \, \bar{\xi}_L\bar{\psi}_{-I'R} + \sqrt{2} \, \bar{\xi}_R\bar{\psi}_{-I'L}, \nn \\
\delta_\xi \bar{\psi}_{-I'R} & = & i2\sqrt{2} \, \xi_L\cD_R\phi_{-I'}^\dagger -\sqrt{2} \, \xi_R(\phi-\twm_{-I'})\phi_{-I'}^\dagger 
+\sqrt{2} \, \bar{\xi}_R F_{-I'}^\dagger, \nn \\ 
\delta_\xi \bar{\psi}_{-I'L} & = & -i2\sqrt{2} \, \xi_R\cD_L\phi_{-I'}^\dagger +\sqrt{2} \, \xi_L(\bar{\phi}-\twm_{-I'}^*)\phi_{-I'}^\dagger 
+\sqrt{2} \, \bar{\xi}_L F_{-I'}^\dagger, \nn \\ 
\delta_\xi F_{-I'}^\dagger & = & -i2\sqrt{2} \, \xi_L\cD_R\bar{\psi}_{-I'L} -i2\sqrt{2} \, \xi_R\cD_L\bar{\psi}_{-I'R}  \nn \\
 & & +\sqrt{2} \, \xi_R(\phi-\twm_{-I'})\bar{\psi}_{-I'L} +\sqrt{2} \, \xi_L(\bar{\phi}-\twm_{-I'}^*)\bar{\psi}_{-I'R} \nn \\
 & & +2i\xi_L\lambda_R\phi_{-I'}^\dagger -2i\xi_R\lambda_L\phi_{-I'}^\dagger. 
\label{2dsusy_mat-_twm} 
\eea

\section{On Lattice Perturbation for ${\rm U}(1)_A$ Anomaly}
\label{app:lat_pert}
\setcounter{equation}{0}
In this appendix, we present the explicit form of the ${\rm U}(1)_A$-Noether current in the SYM sector of the lattice theory 
for the completeness. 
Also, the form of the interaction terms in the matter sector used to the perturbative computation of the ${\rm U}(1)_A$ anomaly 
is expressed up to the relevant orders. 

\subsection{${\rm U}(1)_A$ Noether Current in Lattice SYM}
${\rm U}(1)_A$-Noether current derived from the SYM part (\ref{S_lat_SYM1'}) is explicitly given as 
\bea
J^{\rm SYM}_{A0}(x) & = & \frac{a^2}{g_0^2}\left\{\frac{i}{1-\frac{1}{\epsilon^2}||1-U_{01}(x)||^2} \right. \nn \\
 & &\hspace{1cm} \times \tr \left[\chi(x)\left(U_0(x)\psi_1(x+\hat{0})U_1(x+\hat{0})U_0(x+\hat{1})^{-1}U_1(x)^{-1} \right.\right. \nn \\
& & \hspace{2cm} \left.\left.  +U_1(x)U_0(x+\hat{1})U_1(x+\hat{0})^{-1}\psi_1(x+\hat{0})U_0(x)^{-1}\right)\right]   \nn \\
 & &\hspace{0.5cm}  +\frac{1}{\left(1-\frac{1}{\epsilon^2}||1-U_{01}(x)||^2\right)^2} \, \frac{1}{\epsilon^2} \, \tr \left(\chi(x)\Phi(x)\right) \nn \\
 & &\hspace{1cm}  \times \tr \left[\psi_1(x+\hat{0})\left(U_1(x+\hat{0})U_0(x+\hat{1})^{-1}U_1(x)^{-1}U_0(x) \right.\right. \nn \\
 & & \hspace{2cm} \left.\left.  -U_0(x)^{-1}U_1(x)U_0(x+\hat{1})U_1(x+\hat{0})^{-1}\right)\right] \nn \\
  & & \hspace{0.5cm} +\tr\left[2\phi(x)D_0\bar{\phi}(x) -2\bar{\phi}(x)D_0\phi(x) -i\psi_0(x)U_0(x)\eta(x+\hat{0})U_0(x)^{-1} \right. \nn \\
 & & \hspace{1cm} \left.\left. +2a\psi_0(x)\psi_0(x)U_0(x)\bar{\phi}(x+\hat{0})U_0(x)^{-1}\right] \frac{}{} \right\},  
\\
 J^{\rm SYM}_{A1}(x) & = & \frac{a^2}{g_0^2}\left\{\frac{-i}{1-\frac{1}{\epsilon^2}||1-U_{01}(x)||^2} \right. \nn \\
  & &\hspace{1cm} \times \tr \left[\chi(x)\left(U_0(x)U_1(x+\hat{0})U_0(x+\hat{1})^{-1}\psi_0(x+\hat{1}) U_1(x)^{-1} \right.\right. \nn \\
& & \hspace{2cm} \left.\left.  +U_1(x)\psi_0(x+\hat{1})U_0(x+\hat{1})U_1(x+\hat{0})^{-1}U_0(x)^{-1}\right)\right]   \nn \\
 & &\hspace{0.5cm}  +\frac{1}{\left(1-\frac{1}{\epsilon^2}||1-U_{01}(x)||^2\right)^2} \, \frac{1}{\epsilon^2} \, \tr \left(\chi(x)\Phi(x)\right) \nn \\
 & &\hspace{1cm}  \times \tr \left[\psi_0(x+\hat{1})\left(U_0(x+\hat{1})U_1(x+\hat{0})^{-1}U_0(x)^{-1}U_1(x) \right.\right. \nn \\
 & & \hspace{2cm} \left.\left.  -U_1(x)^{-1}U_0(x)U_1(x+\hat{0})U_0(x+\hat{1})^{-1}\right)\right] \nn \\
   & & \hspace{0.5cm} +\tr\left[2\phi(x)D_1\bar{\phi}(x) -2\bar{\phi}(x)D_1\phi(x) -i\psi_1(x)U_1(x)\eta(x+\hat{1})U_1(x)^{-1} \right. \nn \\
 & & \hspace{1cm} \left.\left. +2a\psi_1(x)\psi_1(x)U_1(x)\bar{\phi}(x+\hat{1})U_1(x)^{-1}\right] \frac{}{} \right\}
 \eea 
with 
\bea
D_\mu\phi(x) & \equiv & \frac{1}{a}\left(U_\mu(x)\phi(x+\hat{\mu})U_\mu(x)^{-1}-\phi(x)\right), \nn \\
D_\mu\bar{\phi}(x) & \equiv &  \frac{1}{a}\left(U_\mu(x)\bar{\phi}(x+\hat{\mu})U_\mu(x)^{-1}-\bar{\phi}(x)\right). 
\eea

\subsection{Interaction Terms in Matter Sector}
We explicitly write the interaction terms in the matter sector by dividing into the seven parts:
\beq
S^{\rm LAT-int}_{{\rm mat}, \twm} = V^{\rm int}_1 + \cdots + V^{\rm int}_7. 
\eeq
In the calculation of the ${\rm U}(1)_A$ anomaly from the matter multiplets, 
the SYM fields are treated as the external fields. 
Also, in the lattice perturbation, the link variables $U_{\mu}(x)=e^{iaA_{\mu}(x)}$ are expanded with respect to $A_{\mu}(x)$,  
and the external momenta carried by the SYM fields are treated as quantities much smaller than $1/a$.  
For the computation, it is sufficient to keep the terms with 
\beq
(\text{the power of external momenta}) + (\text{the number of the SYM fields}) \leq 2. 
\label{relevant_order}
\eeq
The interaction terms satisfying (\ref{relevant_order}) are as follows. \\
$V^{\rm int}_1$ consists of the three-point gauge-squark couplings (See footnote~\ref{foot:quark} in page~\pageref{foot:quark}.): 
\bea
V^{\rm int}_1 & = & \sum_{I=1}^n \int\frac{\dd^2q}{(2\pi)^2}\frac{\dd^2k}{(2\pi)^2} \left[
\sum_{\mu=0}^1 \left\{2\bq_{\mu}\cos(aq_{\mu}) +k_{\mu} \cos^2(aq_\mu) -iak_{\mu} \bq_{\mu}e^{-iaq_{\mu}} \frac{}{} \right.\right. \nn \\
 & & \left.\hspace{0.5cm} -i\sum_{\nu=0}^1\epsilon_{\mu\nu}k_{\nu}\cos(aq_0)\cos(aq_1) 
 +(ra)^2\bq_{\mu}\hat{q}^2 +\frac{(ra)^2}{2}k_{\mu}e^{-iaq_{\mu}}\hat{q}^2 +(ra)^2\bq_{\mu} \, k\cdot \bq\right\} \nn \\
 & & \hspace{0.3cm}\times \wt{\phi^\dagger}_{+I}(-k-q)\wt{A}_{\mu}(k)\wt{\phi}_{+I}(q) \nn \\
  & & +\sum_{\mu=0}^1 \left\{2\bq_{\mu}\cos(aq_{\mu}) +k_{\mu} \cos^2(aq_\mu) -iak_{\mu} \bq_{\mu}e^{-iaq_{\mu}} \frac{}{} \right. \nn \\
 & & \left.\hspace{0.5cm}  +i\sum_{\nu=0}^1\epsilon_{\mu\nu}k_{\nu}\cos(aq_0)\cos(aq_1) 
 +(ra)^2\bq_{\mu}\hat{q}^2 +\frac{(ra)^2}{2}k_{\mu}e^{-iaq_{\mu}}\hat{q}^2 +(ra)^2\bq_{\mu} \, k\cdot \bq\right\} \nn \\
 & & \hspace{0.3cm}\times \wt{\phi}_{-I}(-k-q)\wt{A}_{\mu}(k) \wt{\phi^\dagger}_{-I}(q) \nn \\
 & & +ra (-\bq_0\cos(aq_1) +i\bq_1\cos (aq_0))\wt{\phi^\dagger}_{+I}(-k-q)
\left(k_0\wt{A}_1(k)-k_1\wt{A}_0(k)\right)\wt{\phi^\dagger}_{-I}(q)
 \nn \\
  & &\left.  +ra (\bq_0\cos(aq_1) +i\bq_1\cos (aq_0))\wt{\phi}_{-I}(-k-q)
\left(k_0\wt{A}_1(k)-k_1\wt{A}_0(k)\right)\wt{\phi}_{+I}(q)  
\frac{}{}\right] \nn \\
& & 
\eea
with $k\cdot\bar{q} \equiv \sum_{\mu=0}^1 k_{\mu}\bar{q}_{\mu}$. \\
$V^{\rm int}_2$ consists of the four-point gauge-squark interactions:  
\bea
V^{\rm int}_2 & = & \sum_{I=1}^n \int\frac{\dd^2q}{(2\pi)^2}\frac{\dd^2k}{(2\pi)^2} \frac{\dd^2\ell}{(2\pi)^2}\left[
\wt{\phi^\dagger}_{+I}(q) \left\{ \sum_{\mu=0}^1 \cos^2(aq_{\mu}) \wt{A}_{\mu}(k)\wt{A}_{\mu}(\ell) \right.\right.  \nn \\
 & & \left. +i\cos(aq_0)\cos(aq_1)[\wt{A}_0(k), \wt{A}_1(\ell)] 
+ (ra)^2\sum_{\mu, \nu=0}^1 \bq_{\mu}\bq_{\nu}\wt{A}_{\mu}(k)\wt{A}_{\nu}(\ell)\right\} \wt{\phi}_{+I}(-k-\ell-q) \nn \\
 & & \hspace{-0.3cm} +\wt{\phi}_{-I}(q)\left\{ \sum_{\mu=0}^1\cos^2(aq_{\mu})\wt{A}_{\mu}(k)\wt{A}_{\mu}(\ell) \right. \nn \\
 & & \left. -i\cos(aq_0)\cos(aq_1)[\wt{A}_0(k), \wt{A}_1(\ell)] 
+ (ra)^2\sum_{\mu, \nu=0}^1 \bq_{\mu}\bq_{\nu}\wt{A}_{\mu}(k)\wt{A}_{\nu}(\ell)\right\} \wt{\phi^\dagger}_{-I}(-k-\ell-q) \nn \\
 & & \hspace{-0.3cm} +\wt{\phi^\dagger}_{+I}(q) \sum_{\mu=0}^1 ra \bq_{\mu}\left\{ -i\cos(aq_0)  [\wt{A}_0(k), \wt{A}_{\mu}(\ell)] 
 -\cos(aq_1) [\wt{A}_1(k), \wt{A}_{\mu}(\ell)] \right\} \nn \\
 & & \hspace{0.2cm} \times  \wt{\phi^\dagger}_{-I}(-k-\ell-q) \nn \\
  & & \hspace{-0.3cm} +\wt{\phi}_{-I}(q)  \sum_{\mu=0}^1 ra \bq_{\mu} \left\{ -i\cos(aq_0) [\wt{A}_0(k), \wt{A}_{\mu}(\ell)] 
 +\cos(aq_1)  [\wt{A}_1(k), \wt{A}_{\mu}(\ell)] \right\} \nn \\
 & &\left.  \hspace{0.2cm} \times  \wt{\phi}_{+I}(-k-\ell-q) \frac{}{} \right].
\eea
$V^{\rm int}_3$ consists of the three-point gauge-quark couplings:
\bea
V^{\rm int}_3 & = & \sum_{I=1}^n \int\frac{\dd^2q}{(2\pi)^2}\frac{\dd^2k}{(2\pi)^2} \left[
\left\{ i\cos(aq_0)+\frac12 ak_0 e^{-iaq_0}\right\} \wt{\bar{\Psi}}_I(q) \wt{A}_0(k) \wt{\Psi}_I(-k-q) \right. \nn \\
 & & \hspace{3cm} + \left\{ -\cos(aq_1) +i\frac12 ak_1 e^{-iaq_1}\right\} \wt{\bar{\Psi}}_I(q) \Sigma_3 \wt{A}_1(k) \wt{\Psi}_I(-k-q)
\nn \\
 & & \hspace{3cm} \left.  +\sum_{\mu=0}^1 ra \left(\bq_{\mu} +\frac12 k_{\mu} e^{-iaq_{\mu}}\right) 
\wt{\bar{\Psi}}_I(q) \wt{\Sigma}_1\wt{A}_{\mu}(k) \wt{\Psi}_I(-k-q) \right]
\eea
with 
\beq
\Sigma_3\equiv \left(\begin{array}{cc} \sigma_3 & 0 \\ 0 & \sigma_3 \end{array} \right), \qquad 
\wt{\Sigma}_1 \equiv \left(\begin{array}{cc} \sigma_1 & 0 \\ 0 & -\sigma_1 \end{array} \right).
\eeq
$V^{\rm int}_4$ contains the three- and four-point interactions of the squarks to the Higgs, $F_{01}$ or $D$: 
\bea
V^{\rm int}_4 & = & \sum_{I=1}^n \int\frac{\dd^2q}{(2\pi)^2}\frac{\dd^2k}{(2\pi)^2} \frac{\dd^2\ell}{(2\pi)^2} 
\left[\frac12 \wt{\phi^\dagger}_{+I}(q)\{\wt{\phi}(k), \wt{\bar{\phi}}(\ell)\} \wt{\phi}_{+I}(-k-\ell-q) \right. \nn \\
 & & \left.\hspace{4.3cm}  +\frac12 \wt{\phi}_{-I}(q)\{\wt{\phi}(k), \wt{\bar{\phi}}(\ell)\} \wt{\phi^\dagger}_{-I}(-k-\ell-q) \right] \nn \\
  & & \hspace{-0.4cm} + \sum_{I=1}^n \int\frac{\dd^2q}{(2\pi)^2}\frac{\dd^2k}{(2\pi)^2} \left[ 
  \wt{\phi^\dagger}_{+I}(q) \left(-\wt{F}_{01}(k)-\wt{D}(k)-\twm_I\wt{\bar{\phi}}(k) -\twm_{+I}^* \wt{\phi}(k) \right) 
  \wt{\phi}_{+I}(-k-q) \right. \nn \\
  & & \left. \hspace{3.2cm} + \wt{\phi}_{-I}(q) \left(\wt{F}_{01}(k)+\wt{D}(k)-\twm_I\wt{\bar{\phi}}(k) -\twm_{-I}^* \wt{\phi}(k) \right) 
  \wt{\phi^\dagger}_{-I}(-k-q) \right]. \nn \\
  & & 
\eea 
$V^{\rm int}_5$ consists of the Yukawa couplings of the Higgs to the quarks:
\bea
V^{\rm int}_5 & = & \sum_{I=1}^n \int\frac{\dd^2q}{(2\pi)^2}\frac{\dd^2k}{(2\pi)^2} \, 
\wt{\bar{\Psi}}_I(q) \left[\begin{array}{cccc} & & & \wt{\phi}(k) \\ & & \wt{\phi}(k) & \\ 
 & \wt{\bar{\phi}}(k) & & \\ \wt{\bar{\phi}}(k) & & & \end{array} \right] \wt{\Psi}_I(-k-q).
\eea
$V^{\rm int}_6$ consists of the Yukawa interactions containing $\chi, \eta$:  
\bea
V^{\rm int}_6 & = & \sum_{I=1}^n \int\frac{\dd^2q}{(2\pi)^2}\frac{\dd^2k}{(2\pi)^2} \left[ 
\wt{\phi^\dagger}_{+I}(q) \left(i\wt{\chi}(k) -\frac12\wt{\eta}(k)\right) \wt{\psi}_{+IL}(-k-q) \right. \nn \\
 & & \hspace{3.2cm}+\wt{\bar{\psi}}_{+IR}(q)\left(-i\wt{\chi}(k) -\frac12\wt{\eta}(k)\right) \wt{\phi}_{+I}(-k-q) \nn \\
 & &  \hspace{3.2cm}+\wt{\phi}_{-I}(q) \left(-i\wt{\chi}(k) -\frac12\wt{\eta}(k)\right) \wt{\bar{\psi}}_{-IR}(-k-q)  \nn \\
 & &\left. \hspace{3.2cm}+\wt{\psi}_{-IL}(q)\left(i\wt{\chi}(k) -\frac12\wt{\eta}(k)\right) \wt{\phi^\dagger}_{-I}(-k-q) \right]. 
\eea
$V^{\rm int}_7$ consists of the Yukawa interactions containing $\psi_{\mu}$: 
\bea
V^{\rm int}_7 & = & \sum_{I=1}^n \int\frac{\dd^2q}{(2\pi)^2}\frac{\dd^2k}{(2\pi)^2} \left[ 
\left\{-i\cos(aq_0) -\frac12 a k_0 e^{-iaq_0}\right\} \left(\wt{\bar{\psi}}_{+IL}(q) \wt{\psi}_0(k) \wt{\phi}_{+I}(-k-q) \right.\right. \nn \\
 & & \hspace{3.5cm} +\wt{\phi^\dagger}_{+I}(q) \wt{\psi}_0(k) \wt{\psi}_{+IR}(-k-q) 
      +\wt{\phi}_{-I}(q) \wt{\psi}_0(k) \wt{\bar{\psi}}_{-IL}(-k-q) \nn \\
  & & \hspace{3.5cm} \left. +\wt{\psi}_{-IR}(q) \wt{\psi}_0(k) \wt{\phi^\dagger}_{-I}(-k-q) \right)\nn \\
 & & \hspace{3cm}+\left\{ \cos (aq_1) -i\frac12 ak_1 e^{-iaq_1}\right\} \left(\wt{\bar{\psi}}_{+IL}(q) \wt{\psi}_1(k) \wt{\phi}_{+I}(-k-q) \right. \nn \\
 & & \hspace{3.5cm} -\wt{\phi^\dagger}_{+I}(q) \wt{\psi}_1(k) \wt{\psi}_{+IR}(-k-q) 
       +\wt{\phi}_{-I}(q) \wt{\psi}_1(k) \wt{\bar{\psi}}_{-IL}(-k-q) \nn \\
 & & \hspace{3.5cm} \left. -\wt{\psi}_{-IR}(q) \wt{\psi}_1(k) \wt{\phi^\dagger}_{-I}(-k-q) \right)\nn \\
 & &\hspace{3cm} +\sum_{\mu=0}^1 ra \left(-\bq_{\mu} -\frac12k_{\mu} e^{-iaq_{\mu}}\right) 
\left(\wt{\bar{\psi}}_{+IL}(q) \wt{\psi}_{\mu}(k) \wt{\phi^\dagger}_{-I}(-k-q) \right. \nn \\
 & & \hspace{3.5cm} -\wt{\phi}_{-I}(q) \wt{\psi}_{\mu}(k)\wt{\psi}_{+IR}(-k-q) -\wt{\phi^\dagger}_{+I}(q) \wt{\psi}_{\mu}(k) 
\wt{\bar{\psi}}_{-IL}(-k-q) \nn \\
 & & \hspace{3.5cm} \left.\left. +\wt{\psi}_{-IR}(q) \wt{\psi}_{\mu}(k) \wt{\phi}_{+I}(-k-q)\right) \frac{}{} \right]. 
\eea


\end{document}